\documentclass[journal=jacsat, manuscript=article, layout=onecolumn]{achemso}

\makeatletter
\def\@fnsymbol#1{\ensuremath{\ifcase#1\or \ddagger\or *\or \dagger\or	\mathsection\or \mathparagraph\or **\or \dagger\dagger\or \ddagger\ddagger \else\@ctrerr\fi}}
\makeatother

\usepackage{graphicx}
\usepackage{color,amsmath,amssymb}
\usepackage[normalem]{ulem}
\usepackage{hyperref}
\usepackage{lipsum}

\newcommand{\Bpar}{B_{\parallel}}
\newcommand{\Lsc}{L_{\mathrm{SC}}}

\newcommand{\Vglob}{V_{\mathrm{Global}}}
\newcommand{\Bperp}{B_{\perp}}
\newcommand{\Iref}{I_{\mathrm{ref.}}}
\newcommand{\Vtg}{V_{\mathrm{TG}}}
\newcommand{\dI}{\Delta I}
\newcommand{\avedI}{\langle\Delta I/2\rangle}
\newcommand{\Bphi}{B^{\mathit{\Phi}}_{\parallel}}
\newcommand{\Vsd}{V_{\mathrm{SD}}}
\newcommand{\Vt}{V_{\mathrm{T}}}
\newcommand{\Bperiod}{B_{\mathrm{Period}}}
\newcommand{\Bt}{B_{\mathrm{t}}}
\newcommand{\Vprobe}{V_{\mathrm{Probe}}}
\newcommand{\Idc}{I_{\mathrm{DC}}}
\newcommand{\Vtwo}{V_{2}}
\newcommand{\Vac}{V_{\mathrm{AC}}}
\newcommand{\Ione}{I_{1}}
\newcommand{\Vone}{V_{1}}
\newcommand{\Ea}{E_{\mathrm{A}}}
\newcommand{\VtL}{V_{\mathrm{T,L}}}
\newcommand{\VtR}{V_{\mathrm{T,R}}}

\newcommand{\Bzero}{B_{0}}
\newcommand{\Izero}{I_{0}}
\newcommand{\Ial}{I_{\mathrm{Al}}}
\newcommand{\Wcons}{W_{\mathrm{cons.}}}

\title{Zeeman and Orbital Driven Phase Transitions in Planar Josephson Junctions}

\author{D. Z. Haxell}
\affiliation{IBM Research Europe - Zurich, 8803 R\"uschlikon, Switzerland}

\author{M. Coraiola}
\affiliation{IBM Research Europe - Zurich, 8803 R\"uschlikon, Switzerland}

\author{D. Sabonis}
\affiliation{IBM Research Europe - Zurich, 8803 R\"uschlikon, Switzerland}

\author{M. Hinderling}
\affiliation{IBM Research Europe - Zurich, 8803 R\"uschlikon, Switzerland}

\author{S. C. ten Kate}
\affiliation{IBM Research Europe - Zurich, 8803 R\"uschlikon, Switzerland}

\author{E. Cheah}
\affiliation{Solid State Laboratory, ETH Z\"urich, 8093 Z\"urich, Switzerland}

\author{F. Krizek}
\affiliation{IBM Research Europe - Zurich, 8803 R\"uschlikon, Switzerland}
\alsoaffiliation{Solid State Laboratory, ETH Z\"urich, 8093 Z\"urich, Switzerland}
\alsoaffiliation[Present address: ]{Institute of Physics, Czech Academy of Sciences, 162 00 Prague, Czech Republic}

\author{R. Schott}
\affiliation{Solid State Laboratory, ETH Z\"urich, 8093 Z\"urich, Switzerland}

\author{W. Wegscheider}
\affiliation{Solid State Laboratory, ETH Z\"urich, 8093 Z\"urich, Switzerland}

\author{F. Nichele}
\email{fni@zurich.ibm.com}
\affiliation{IBM Research Europe - Zurich, 8803 R\"uschlikon, Switzerland}

\begin{document}
\date{\today}

\begin{abstract}
We perform supercurrent and tunneling spectroscopy measurements on gate-tunable InAs/Al Josephson junctions (JJs) in an in-plane magnetic field, and report on phase shifts in the current-phase relation measured with respect to an absolute phase reference. The impact of orbital effects is investigated by studying multiple devices with different superconducting lead sizes. 
At low fields, we observe gate-dependent phase shifts of up to ${\varphi_{0}=0.5\pi}$ which are consistent with a Zeeman field coupling to highly-transmissive Andreev bound states via Rashba spin-orbit interaction.
A distinct phase shift emerges at larger fields, concomitant with a switching current minimum and the closing and reopening of the superconducting gap. These signatures of an induced phase transition, which might resemble a topological transition, scale with the superconducting lead size, demonstrating the crucial role of orbital effects.
Our results elucidate the interplay of Zeeman, spin-orbit and orbital effects in InAs/Al JJs, giving new understanding to phase transitions in hybrid JJs and their applications in quantum computing and superconducting electronics.
\end{abstract}

\maketitle

Josephson junctions (JJs) defined in hybrid superconductor-semiconductor materials are the subject of intense investigation as building blocks of gate-tunable superconducting~\cite{deLange2015,Larsen2015,Casparis2018,Kringhoj2021,Hertel2022} and Andreev~\cite{Andreev1964,Beenakker1991,Furusaki1991,Desposito2001,Zazunov2003,Chtchelkatchev2003,Padurariu2010,Janvier2015,Hays2018,Tosi2019,Hays2021,Matute2022,PitaVidal2023} qubits, along with transistors~\cite{Gheewala1980,Clark1980,Kleinsasser1989,Wen2021}, mixers~\cite{Leroux2022} and rectifiers~\cite{Suoto2022} for superconducting electronics. Additional functionalities are enabled by the interplay between spin-orbit interaction and external magnetic fields, including spin-dependent~\cite{Linder2015,Hart2017} and non-reciprocal supercurrents~\cite{Baumgartner2022a,Costa2022,Baumgartner2022b}, topological phase transitions~\cite{Pientka2017,Hell2017,Fornieri2019,Ren2019,Dartiailh2021} and anomalous shifts in the ground state~\cite{Konschelle2015,Szombati2016,Murani2017,Spanton2017,Li2019,Assouline2019,Mayer2020,Haxell2023b}.
The latter constitute a shift in the energy minimum away from a phase difference $\varphi=0$ across the JJ, to $0<\varphi<\pi$ by breaking of time-reversal symmetry~\cite{Bezuglyi2002,Buzdin2008,Liu2010,Yokoyama2014,Bergeret2015} or to $\varphi=\pi$ by a Zeeman-induced phase transition~\cite{Fulde1964,Larkin1964,Yokoyama2014}.

Epitaxially-grown InAs/Al heterostructures~\cite{Shabani2016,Cheah2023} are a promising platform to realize these complex devices, due to their high electron mobility, excellent superconducting properties~\cite{Kjaergaard2017,Nichele2020} and prospect of scalability. To date, tunneling spectroscopy experiments of planar InAs/Al JJs have revealed the onset of zero-energy states at large in-plane magnetic fields~\cite{Fornieri2019,Ren2019}, and more refined devices~\cite{Banerjee2023} have since shown zero-energy states accompanied by closure and reopening of the superconducting gap, consistent with a topological transition. Supercurrent measurements in superconducting quantum interference devices (SQUIDs) demonstrated gate-tunable phase shifts in small magnetic fields~\cite{Mayer2020}, as well as large phase jumps at larger fields~\cite{Dartiailh2021} accompanied by a minimum in the supercurrent amplitude, also consistent with a topological transition~\cite{Pientka2017}. However, several questions remain on the behavior of planar JJs subject to in-plane magnetic fields. For instance, Ref.~\cite{Mayer2020} reported anomalous phase shifts at small magnetic fields which were considerably larger than theoretical expectations~\cite{Buzdin2008}. Additionally, orbital effects can resemble the behavior expected from a topological transition~\cite{Pientka2017,Hess2023}: a magnetic flux threading the cross-section underneath the superconducting leads can produce non-monotonic switching currents~\cite{Fornieri2019,Drachmann2021} together with closure and reopening of the induced superconducting gap. 
In this context, it is crucial to understand the mechanisms underlying phase shifts in planar JJs in an in-plane magnetic field, to fully harness their properties in quantum computation and superconducting electronics applications.

In this work, we present a comprehensive investigation of planar SQUIDs in in-plane magnetic fields. An advanced device geometry allowed simultaneous measurements of the Andreev bound state (ABS) spectrum of a planar JJ and its current-phase relation (CPR), including anomalous phase shifts relative to an absolute phase reference. The role of orbital effects was studied by measuring several devices with varying size of the superconducting leads. For small in-plane magnetic fields oriented perpendicular to the current flow in the JJ, that is along the direction of the Rashba spin-orbit field, we observed phase shifts in the CPR which depended linearly on magnetic field and varied strongly with gate voltage, similar to Ref.~\cite{Mayer2020}. For simplicity, we define this as a Type~A phase shift. Spectroscopic measurements demonstrated that Type~A phase shifts in the CPR were highly correlated with phase shifts of ballistic ABSs in the JJs, but were found to be independent on the size of the superconducting contacts. Upon further increase in magnetic field, we observed a rapid increase of the anomalous phase shift, which did not depend on gate voltage but was instead strongly correlated with the length of the superconducting contacts, indicating an orbital origin. We define this as a Type~B phase shift. Strikingly, Type~B phase shifts were accompanied by both a local minimum in the amplitude of the CPR and a closure and reopening of the superconducting gap, which might resemble a topological transition. We discuss similarities and differences of our observations with respect to previous work. Our results establish a new baseline understanding of InAs/Al JJs subject to in-plane magnetic fields, and guide towards a more complete understanding of anomalous phase shifts and topological transitions in planar JJs.

\section{Results and Discussion}
\begin{figure*}
	\includegraphics[width=\textwidth]{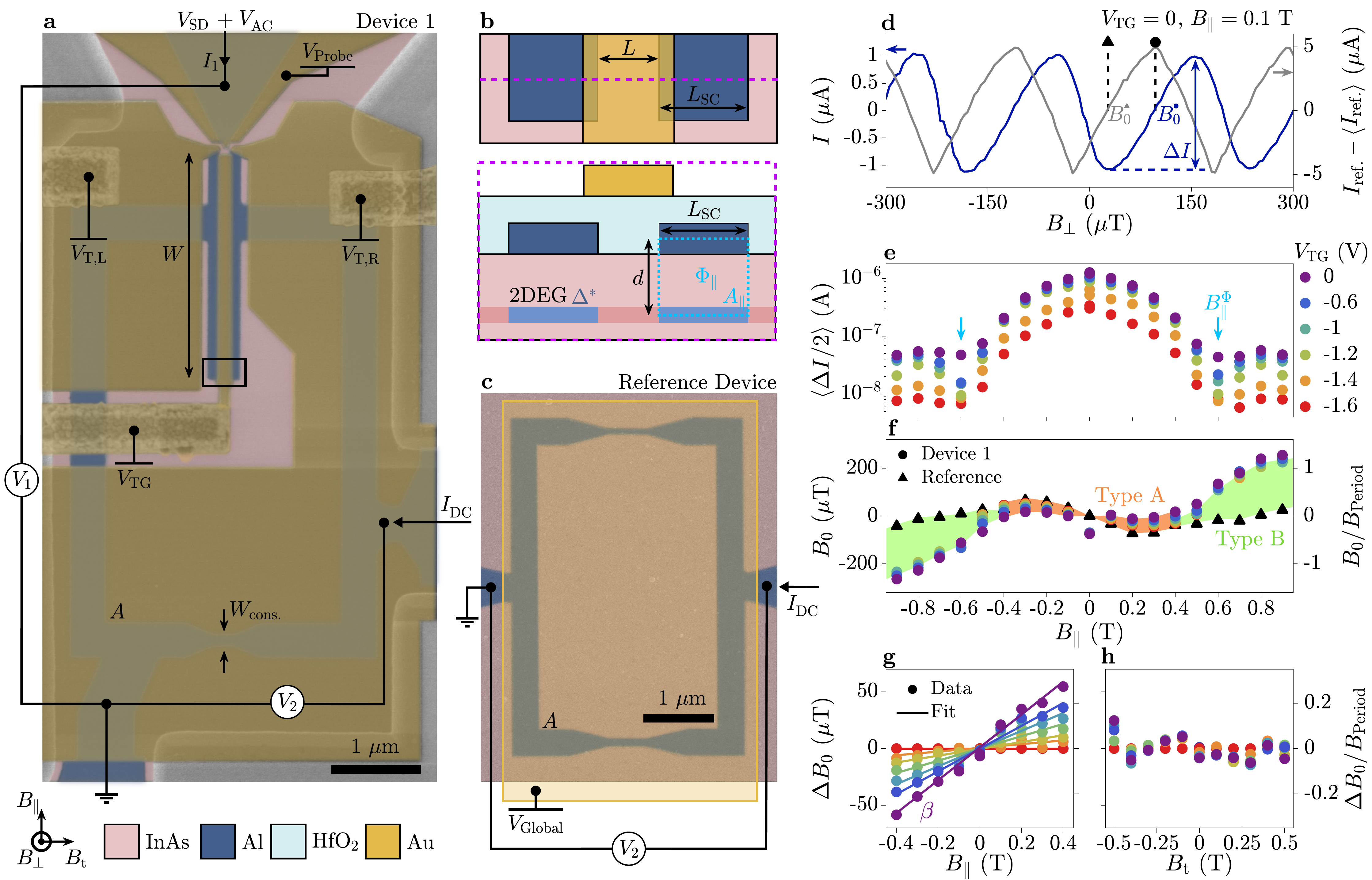}
	\caption{Device under study and current-biased measurements in an in-plane magnetic field $\Bpar$. (a) False-colored scanning electron micrograph (SEM) of Device 1, the planar superconducting quantum interference device (SQUID), consisting of InAs (pink) and Al (blue). Exposed InAs regions were controlled via electrostatic gates (yellow). (b) Schematic zoom-in of the Josephson junction region (top), with junction length $L=80~\mathrm{nm}$ and superconducting lead length $\Lsc=250~\mathrm{nm}$ indicated. The purple dashed line indicates the position of a schematic cross-section (bottom). An in-plane magnetic field $\Bpar$ generates a flux $\mathit{\Phi}_{\parallel}$ between the superconducting leads and the proximitized two-dimensional electron gas (2DEG), with area $A_{\parallel}=\Lsc d$. (c) False-colored SEM of the Reference Device, prior to gate deposition, consisting of two Al constrictions embedded in a superconducting loop. A global gate $\Vglob$ is indicated schematically (yellow). (d) Switching current $I$ of Device 1 as a function of perpendicular magnetic field $\Bperp$ (blue), at a top-gate voltage $\Vtg=0$ and $\Bpar=0.1~\mathrm{T}$, after removing a background of $37~\mathrm{\mu A}$ corresponding to the Al constriction. Switching current of the Reference Device $\Iref$ (grey) at the same $\Bpar$, after subtracting the average $\langle\Iref\rangle$. The zero-current position for Device 1 (Reference Device) is indicated by the circle (triangle). (e) Averaged half-amplitude of a SQUID oscillation $\avedI$ as a function of in-plane magnetic field $\Bpar$, for different top gate voltages $\Vtg$ (colors). A minimum in $\avedI$ occurred at $\Bpar=\Bphi$ (turquoise arrows). (f) Shift in perpendicular magnetic field $\Bzero$ of Device 1 (circles) and Reference Device (triangles), as a function of $\Bpar$. Deviation of Device 1 from the Reference Device is highlighted in orange for $|\Bpar|\lesssim0.4~\mathrm{T}$ and green for $|\Bpar|\gtrsim0.4~\mathrm{T}$. (g) Perpendicular field shift $\Delta\Bzero$ for small $\Bpar$ for each $\Vtg$ (circles), with a linear fit (lines) of gradient $\beta$. Data is plotted relative to $\Vtg=-1.6~\mathrm{V}$. (h) Perpendicular field shift $\Delta\Bzero$ for in-plane fields $\Bt$ applied along the transverse direction.}
	\label{fig1}
\end{figure*}

Experiments were performed on six devices. Figure~\ref{fig1}(a) shows a false-colored scanning electron micrograph of Device~1, the principal device under study, which consisted of a planar SQUID fabricated in a heterostructure of InAs (pink) and epitaxial Al (blue)~\cite{Shabani2016,Cheah2023}. The device was covered by a $\rm{HfO_{2}}$ dielectric layer, onto which Au gate electrodes (yellow) were deposited. The superconducting loop, defined in the epitaxial Al, contained a superconductor-normal semiconductor-superconductor (SNS) JJ and a narrow Al constriction. The SNS junction had length $L=80~\mathrm{nm}$, width $W=2.5~\mathrm{\mu m}$ and Al leads of length $\Lsc=250~\mathrm{nm}$. The constriction had width $W_{\mathrm{cons.}}=130~\mathrm{nm}$, chosen to limit the switching current of the planar SQUID, while still being much larger than that of the SNS junction. This asymmetric configuration resulted in a phase drop across the SNS junction of $\varphi\approx2\pi(\mathit{\Phi}/\mathit{\Phi}_{0})$, where a flux $\mathit{\Phi}=A\Bperp$ threaded the area $A=10.2~\mathrm{(\mu m)^{2}}$ enclosed by the SQUID loop ($\mathit{\Phi}_{0}=h/2\mathrm{e}$ is the superconducting flux quantum). Differently from previous work~\cite{Fornieri2019,Nichele2020,Mayer2020,Dartiailh2021}, where two InAs JJs were used, the Al constriction cannot introduce anomalous phase shifts in an in-plane magnetic field due to the absence of spin-orbit and orbital effects. A superconducting probe was integrated close to one end of the SNS junction, comprising a contact of epitaxial Al separated from the SNS junction by a tunnel barrier defined in the InAs. The transparency of the tunnel barrier was controlled by the gate voltages $\VtL$ and $\VtR$, applied to the left and right tunnel gates respectively. The carrier density in the SNS junction was controlled via a top-gate voltage $\Vtg$. An additional gate was kept at $\Vprobe=0$ throughout. Devices~2 to 5 were similar to Device~1 except for $\Lsc$, resulting in different orbital coupling to in-plane magnetic fields [see Fig.~\ref{fig1}(b)]. Each measurement presented here was acquired in parallel with measurements of a Reference Device fabricated on the same chip, which consisted of a SQUID with two Al constrictions of different widths [see Fig.~\ref{fig1}(c)]. Parallel conduction in the InAs surrounding Reference Devices was prevented by setting a global gate to $\Vglob=-1.5~\mathrm{V}$.

Switching currents $I$ were measured using fast current ramps and voltage triggers. A ramped current $\Idc$ was injected into the SQUID loop while monitoring the voltage $\Vtwo$ across the device with an oscilloscope. The switching current was defined as the value of $\Idc$ at which $\Vtwo$ exceeded a threshold. Particular care was taken to inject the current $\Idc$ by symmetrically biasing the measurement circuit, to prevent significant voltage build-up between SQUID and gates. Each CPR data point shown here was obtained by averaging over 32 data points measured with $\Idc>0$ and 32 with $\Idc<0$. This procedure allowed us to improve the experimental accuracy, limit the effect of the broad switching current distributions typical of planar devices~\cite{Haxell2023} and cancel trivial phase shifts originating from the kinetic inductance of the loop~\cite{Nichele2017}. The CPR of the SNS junction was obtained by subtracting the switching current of the Al constriction $\Ial$ from that of the SQUID loop, which had a value between $30$ and $45~\mathrm{\mu A}$ for all devices. Tunneling conductance measurements were performed by low-frequency lock-in techniques. A voltage bias $\Vsd+\Vac$ was sourced at the tunneling probe and the resulting AC current $\Ione$ and voltage $\Vone$ gave the differential conductance $G\equiv\Ione/\Vone$. Global magnetic fields were applied via a three-axis vector magnet, nominally along the directions $\Bperp$, $\Bpar$ and $\Bt$ as indicated in Fig.~\ref{fig1}(a). Further details on electronic measurements and on the procedures used to accurately align the chip to the external magnetic field are presented in the Supporting Information.

Figure~\ref{fig1}(d) shows the CPR of Device 1 at $\Vtg=0$ (blue line, left axis) and Reference Device (gray line, right axis) at $\Bpar=0.1~\mathrm{T}$. We highlight the maximum switching current $\dI/2$ and a $\Bperp$-field shift $\Bzero$, which was measured where the CPR crossed zero with positive slope (circle and triangle for Device 1 and Reference Device, respectively). Figures~\ref{fig1}(e) and (f) show $\dI/2$ and $\Bzero$, respectively, as a function of $\Bpar$ and for various values of $\Vtg$. Black triangles in Fig.~\ref{fig1}(e) represent magnetic field shifts measured in the Reference Device. In Fig.~\ref{fig1}(e) we plot $\avedI$, that is the maximum supercurrent $\dI/2$ averaged over positive and negative $\Idc$. We observe a non-monotonous dependence of $\avedI$ as a function of $\Bpar$, with minima at $\Bpar=\pm|\Bphi|=\pm0.6~\mathrm{T}$ (see turquoise arrow). The magnetic field shift $\Bzero$ in Fig.~\ref{fig1}(f) shows two distinctive trends. For $|\Bpar|\lesssim0.4~\mathrm{T}~$, $\Bzero$ shows a systematic deviation with respect to the Reference Device (Type~A shift, orange shaded area). Type~A shifts were larger for $\Vtg=0$ (purple) than for $\Vtg=-1.6~\mathrm{V}$ (red). For $|\Bpar|\gtrsim0.4~\mathrm{T}$ we observe a more pronounced shift (Type~B shift, green shading), without any measurable gate voltage dependence. Notably, at $\Bpar=\pm\Bphi$, where the supercurrent was at a minimum, the shift was approximately half a SQUID period, corresponding to a phase shift of $\sim\pm\pi$. At $\Bpar=0.9~\mathrm{T}$, the magnetic field shift accumulated in Device~1 exceeded one SQUID period. Finally, we note a weak "S"-shaped dependence of $\Bzero$, both for Device~1 and the Reference Device, which persisted after accurate alignment of the external magnetic field (see Supporting Information). We speculate that the residual trend in $\Bzero$ originated from flux focusing~\cite{Suominen2017} or a non-linearity of the vector magnet.
Figure~\ref{fig1}(g) shows $\Delta\Bzero$, that is $\Bzero$ as in Fig.~\ref{fig1}(f) after subtraction of the data at $\Vtg=-1.6~\mathrm{V}$, which is the most negative top-gate voltage and follows the trend of the Reference Device for $|\Bpar|\leq0.4~\mathrm{T}$. At each gate voltage, the field shift (circles) was approximately linear in $\Bpar$, as highlighted by the linear fits (solid lines). The slope $\beta$ extracted from the linear fits increased for more positive $\Vtg$. Remarkably, no significant phase shift of either Type~A or B was observed for in-plane fields $\Bt$ applied along the transverse direction, as shown in Fig.~\ref{fig1}(h) for Type~A shifts (see Supporting Information for further details). The lack of Type~A shifts as a function of $\Bt$ implies a direction-dependent coupling to the external field, with a coupling strength indicated by $\beta$.

\begin{figure*}
	\includegraphics[width=\textwidth]{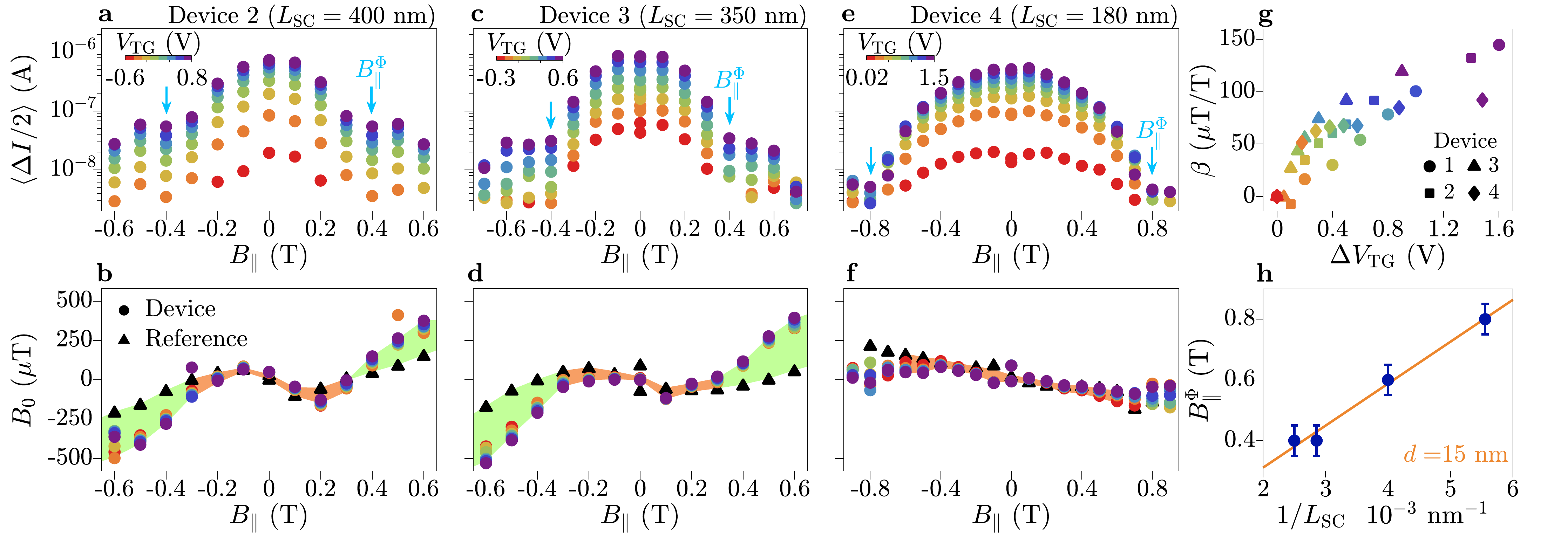}
	\caption{Switching current and perpendicular magnetic field shift for devices with varying $\Lsc$. (a) Average oscillation amplitude $\avedI$ of Device 2: a planar superconducting quantum interference device (SQUID) with a superconducting lead length of $\Lsc=400~\mathrm{nm}$, as a function of in-plane magnetic field $\Bpar$ for different top-gate voltages $\Vtg$ (colors). Minima in the oscillation amplitude, $\Bphi$, are marked with the blue arrows. (b) Shift in perpendicular magnetic field, $\Bzero$, of Device 2 (circles) and the Reference Device (triangles), as a function of $\Bpar$. Deviation of Device 2 from the Reference Device is highlighted in orange for small $\Bpar$ and green for large $\Bpar$. (c,~d) and (e,~f) are the same as (a,~b) for Devices 3 and 4, respectively. All devices are identical in design other than the length of the superconducting contacts, which is $\Lsc=350~\mathrm{nm}$ for Device 3 and $\Lsc=180~\mathrm{nm}$ for Device 4. (g) Gradient $\beta$ of Type~A phase shifts at small $\Bpar$, for Devices 1–4 (circles, squares, triangles and diamonds respectively), plotted against the change in top-gate voltage $\Delta\Vtg$ with respect to the minimum value. (h) In-plane magnetic field where the supercurrent is minimum, $\Bphi$, as a function of inverse superconducting lead length $1/\Lsc$ (blue circles), with a linear fit $\Bphi=(\mathit{\Phi}_{0}/d)/\Lsc$ (orange line) giving $d=15~\mathrm{nm}$.}
	\label{fig2}
\end{figure*}

We now present CPR data obtained from Devices~2, 3 and 4, where $\Lsc$ was $400$, $350$ and $180~\mathrm{nm}$, respectively. Switching currents $\dI/2$ are shown in Figs.~\ref{fig2}(a,~c,~e) for Devices~2-4 respectively, with field shifts $\Bzero$ in Figs.~\ref{fig2}(b,~d,~f) for each device (colored markers) alongside those of a Reference Device measured in parallel (black triangles). Devices~2, 3 and 4 showed a qualitatively similar behavior to Device~1, despite having $\Bphi=0.4~\mathrm{T}$, $\Bphi=0.4~\mathrm{T}$ and $\Bphi=0.8~\mathrm{T}$, respectively. We repeated the analysis on Type~A phase shifts presented in Fig.~\ref{fig1}(g) on the data of Fig.~\ref{fig2}(b,~d,~f), and show the extracted $\beta$ in Fig.~\ref{fig2}(g) [see Supporting Information for more details]. As each device operated in a different range of $\Vtg$, we compare them by plotting $\beta$ as a function of $\Delta\Vtg$, the top-gate voltage relative to the most negative value at which oscillations were observed. Despite some scattering for small $\Delta\Vtg$, where data analysis is intricate due to the small switching current, we note that $\beta$ follows a similar trend for all devices. In particular, $\beta$ increases with $\Delta\Vtg$ and does not depend on $\Lsc$. Figure~\ref{fig2}(h) shows $\Bphi$ as a function of the inverse superconducting lead length $1/\Lsc$. The data (blue circles) followed a linear trend, fitted by $\Bphi=(\mathit{\Phi}_{0}/d)/\Lsc$ (orange line) describing one flux quantum threading an area $\Lsc d$. The result of $d=15~\mathrm{nm}$ agrees with the separation of Al and InAs layers, indicating a crucial role of orbital effects in inducing Type~B phase shifts.

\begin{figure}
	\includegraphics[width=\columnwidth]{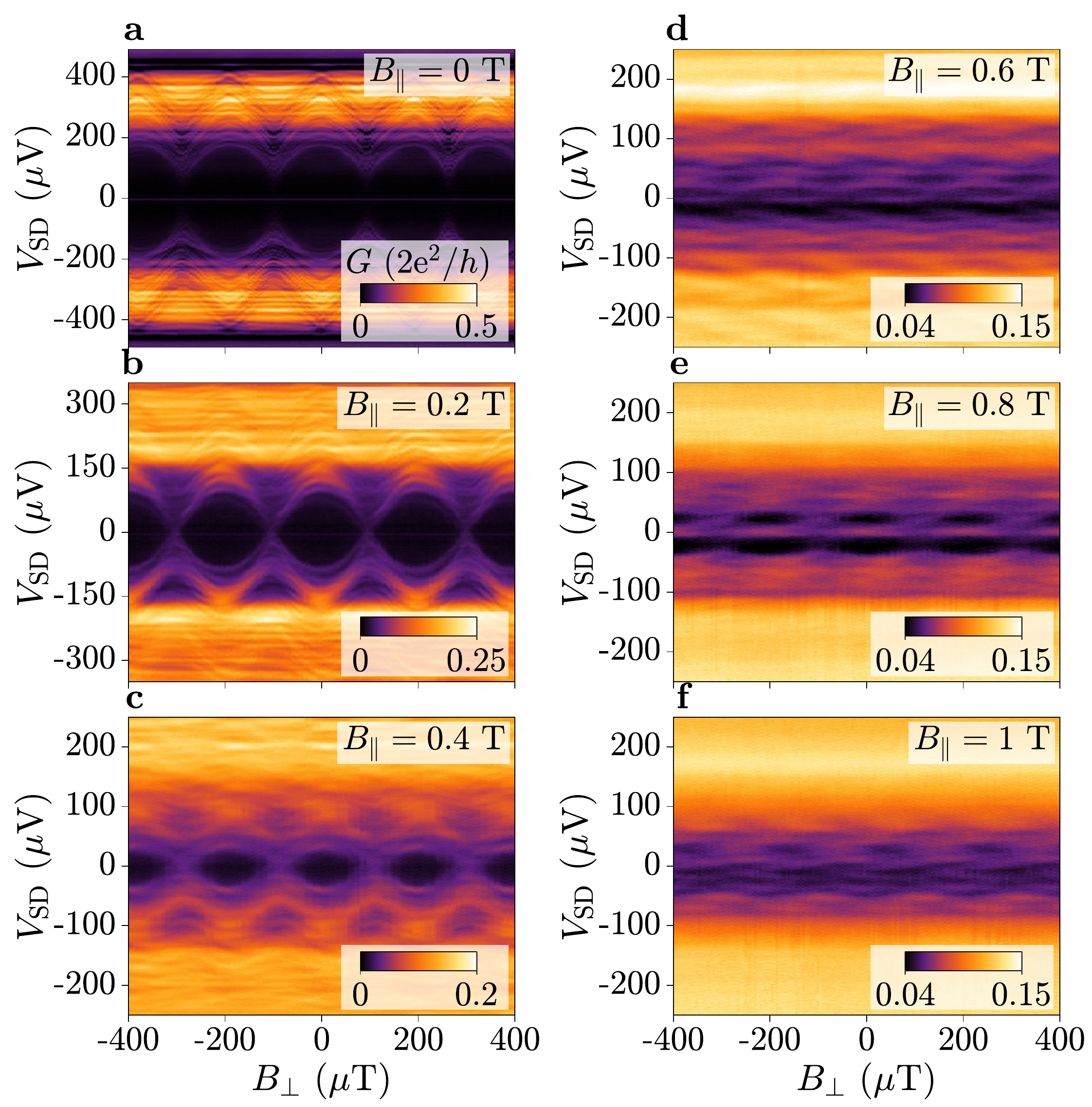}
	\caption{Tunneling spectroscopy of Andreev bound states as a function of in-plane magnetic field $\Bpar$. (a-f) Differential conductance $G$ through the tunneling probe, as a function of source-drain bias voltage $\Vsd$ and perpendicular magnetic field $\Bperp$, for increasing values of $\Bpar$. Measurements were taken at a top-gate voltage of $\Vtg=-1~\mathrm{V}$, with tunnel-barrier voltages $(\VtL,\VtR) = (-1.495,-1.65)~\mathrm{V}$.}
	\label{fig3}
\end{figure}

We now complement CPR measurements with spectroscopic data obtained on Device~1. Figure~\ref{fig3} presents a series of differential conductance maps as a function of $\Bperp$ and $\Vsd$, for increasing values of $\Bpar$. All data were obtained at $\Vtg=-1~\mathrm{V}$ (data at more values of $\Vtg$ are reported in the Supporting Information). As the tunneling probe was constituted by a superconducting lead, the differential conductance $G$ at $\Bpar=0$ indicates the density of states in the junction up to a bias shift of $\pm \mathrm{e}\Delta$. Further conductance peaks at zero and high bias are attributed to a residual supercurrent and multiple Andreev reflection through the tunneling probe, respectively. For $\Bpar\leq 0.2~\mathrm{T}$, the conductance demonstrates a conventional spectrum containing multiple Andreev bound states, some of which have transmission approaching unity and an induced superconducting gap of approximately $180~\mathrm{\mu eV}$. For $\Bpar\geq0.2~\mathrm{T}$, a finite density of states at the Fermi level was induced in the lead facing the tunneling probe, resulting in a direct mapping of the density of states in the junction~\cite{Suominen2017}. For $\Bpar=0.4~\mathrm{T}$, phase-dependent conductance features approached zero energy, resulting in a significant decrease of the superconducting gap [Fig.~\ref{fig3}(c)]. For $\Bpar=\Bphi=0.6~\mathrm{T}$ [Fig.~\ref{fig3}(d)], conductance features oscillated close to $\Vsd=0$ with no clear separation between states at positive and negative bias. As $\Bpar$ was further increased, a gap reopened in the Andreev bound state spectrum, with discrete states around zero energy. Finally, the gap closed for $\Bpar\geq1~\mathrm{T}$. Conductance features close to $\Vsd=0$ in Fig.~\ref{fig3}(e) were reminiscent of zero-bias peaks reported for similar devices at high in-plane magnetic fields and understood in terms on topological states~\cite{Fornieri2019,Ren2019}. However, zero-bias features of Fig.~\ref{fig3}(d) were not robust to small changes in the top-gate voltage $\Vtg$ or tunnel gate voltage $\Vt$ (see Supporting Information). 

\begin{figure}
	\includegraphics[width=\columnwidth]{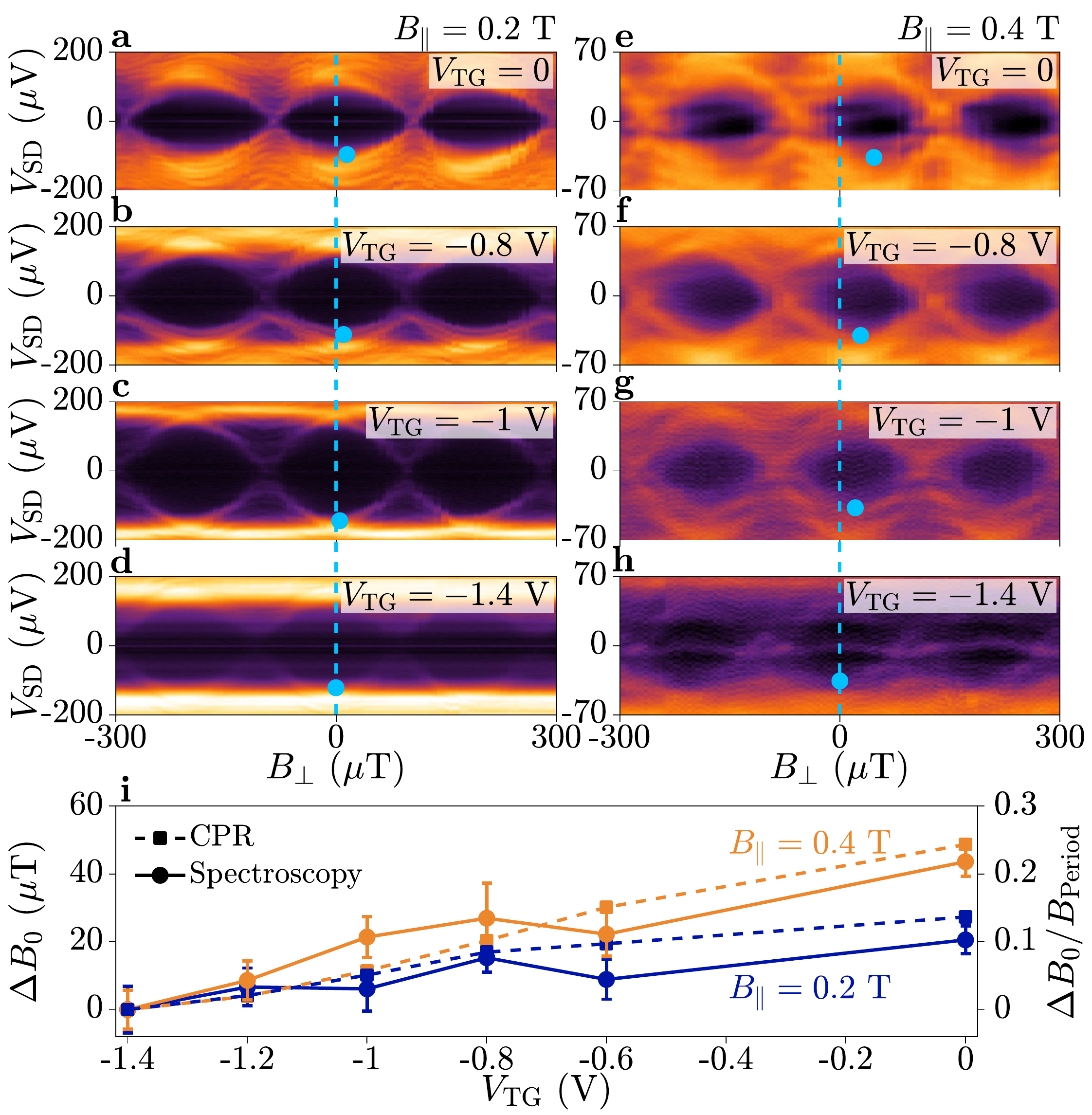}
	\caption{Top-gate dependence of the energy minimum at a finite in-plane magnetic field $\Bpar$. (a-d) Differential conductance $G$ as a function of bias $\Vsd$ and perpendicular magnetic field $\Bperp$, at an in-plane magnetic field of $\Bpar=0.2~\mathrm{T}$. Spectroscopy was performed at a top-gate voltage of $\Vtg=\{0,-0.8,-1,-1.4\}~\mathrm{V}$, respectively. The blue dashed line indicates the energy minimum at $\Vtg=-1.4~\mathrm{V}$. Blue markers show the shift of the energy minimum as a function of $\Vtg$ relative to $\Vtg=-1.4~\mathrm{V}$. (e-h) Bias-dependent spectroscopy as in (a-d) at an in-plane magnetic field of $\Bpar=0.4~\mathrm{T}$. (i) Shift in perpendicular magnetic field $\Delta\Bzero$ relative to $\Vtg=-1.4~\mathrm{V}$, at an in-plane magnetic field of $\Bpar=0.2~\mathrm{T}$ (blue) and $\Bpar=0.4~\mathrm{T}$ (orange), obtained from tunneling spectroscopy (circles, solid lines) and current-phase relation (CPR) measurements (squares, dashed lines). The phase shift $\varphi_{0}/2\pi\equiv\Delta\Bzero/\Bperiod$ is plotted on the right axis.}
	\label{fig4}
\end{figure}

Figure~\ref{fig4} compares spectroscopic maps obtained at $\Bpar=0.2~\mathrm{T}$ (a-d) and $0.4~\mathrm{T}$ (e-h), for multiple values of $\Vtg$. The value of $\Bperp$ at which the ABS energy was closest to the gap was found for each value of $\Vtg$, as indicated by the blue circles. This was determined as the $\Bperp$ value where the gradient $\mathrm{\partial}G/\mathrm{\partial}\Bperp$ was zero, at a fixed bias $\Vsd$ and averaged over multiple periods. Blue dashed lines indicate the minimum energy position at $\Vtg=-1.4~\mathrm{V}$, which is defined as $\Bperp=0$ in Fig.~\ref{fig4}(d). For both $\Bpar=0.2~\mathrm{T}$ and $0.4~\mathrm{T}$, a clear deviation of the ABS spectrum took place as a function of $\Vtg$. The shift in perpendicular field $\Delta\Bzero$ measured from the ABS spectrum is summarized in Fig.~\ref{fig4}(i) as a function of $\Vtg$ for $\Bpar=0.2~\mathrm{T}$ (blue) and $\Bpar=0.4~\mathrm{T}$ (orange). The Type~A shift $\Delta\Bzero$ obtained from the CPR is plotted on the same axis [squares, dashed lines] and shows remarkable agreement.

After demonstrating the occurrence of two types of anomalous phase shifts taking place in hybrid SQUIDs in in-plane magnetic fields, we now discuss their origin. 
Type~A phase shifts, which were approximately linear in $\Bpar$ and depended on $\Vtg$ [Fig.~\ref{fig1}(g)], are associated with spin-orbit-induced anomalous phase shifts~\cite{Bezuglyi2002,Buzdin2008,Liu2010,Yokoyama2014,Bergeret2015}, as recently reported in similar devices~\cite{Mayer2020}. As phase shifts were much more pronounced for in-plane fields aligned perpendicular to the current flow direction ($\Bpar$) than parallel to it ($\Bt$) [Fig.~\ref{fig1}(h)], and were stronger for higher electron density (more positive $\Vtg$~\cite{Wickramasinghe2018}), we conclude that spin-orbit interaction in our samples is predominantly of Rashba type.

Type~A phase shifts reported here, which are of similar magnitude than in Ref.~\cite{Mayer2020}, are considerably larger than theoretical predictions~\cite{Buzdin2008}. Reference~\cite{Mayer2020} proposed that the observed phase offsets could be explained by the contribution of several low-transmission modes. However, here we show that Type~A shifts obtained from the CPR matched those from tunneling spectroscopy [Fig.~\ref{fig4}], where conductance features at both high and low bias showed a phase shift. Since conductance features at low bias correspond to ABSs with high transmission, we conclude that highly transmissive modes participate in the overall phase shift despite their large Fermi velocity. While this result does not resolve the discrepancy between theoretical predictions and experiments~\cite{Mayer2020}, it rules out diffusive modes with small Fermi velocities as the dominant cause of Type~A phase shifts.

Type~B phase shifts were concomitant with a reentrant supercurrrent and a closure and reopening of the superconducting gap, independent of top-gate voltage $\Vtg$. At $\Bpar=\pm\Bphi$, where the supercurrent was at a minimum and the proximitized superconducting gap was suppressed, the phase shift was $\varphi_{0}\approx\pm\pi$. For $|\Bpar|>\Bphi$, a gap reopened in the ABS spectrum and the phase shift increased to above $2\pi$. A phase shift occurring with a supercurrent minimum and gap closure indicates a $0-\pi$ transition at $\Bpar=\Bphi$, where the minimum ABS energy moves from $\varphi\approx0$ to $\varphi\approx\pi$ due to coupling of the magnetic and superconducting orders by Zeeman interaction~\cite{Fulde1964,Larkin1964,Yokoyama2014}. All experimental signatures of Type~B shifts were shown to depend on the length $\Lsc$, consistent with a flux quantum threading an area $\Lsc d$ underneath the superconducting leads. The experimentally obtained value of $d=15~\mathrm{nm}$ agrees with the separation between the Al and InAs layers ($13.4~\mathrm{nm}$), up to some flux penetration into each layer. We therefore conclude that orbital effects strongly contributed to inducing Type~B phase shifts. Type~B shifts were observed for in-plane fields $\Bpar<1~\mathrm{T}$, much lower than the values $B_{0-\pi}\gtrsim9~\mathrm{T}$ expected for InAs/Al heterostructures~\cite{Dartiailh2021}. We explain this by orbital effects, which were responsible for the induced gap reduction, forcing ABSs to move closer in energy. This enabled ABSs to cross even with small Zeeman splitting. Previous work reported similar phase shifts~\cite{Dartiailh2021}, where a $\pi$ jump in the junction phase was accompanied by a minimum in the switching current. However, phase shifts depended on the top-gate voltage, unlike the Type~B shifts reported here. This shows that orbital effects alone are not sufficient to explain the results of Ref.~\cite{Dartiailh2021}.

\section{Conclusions}
In conclusion, measurements of the current phase relation and Andreev bound state spectrum in hybrid quantum interference devices showed phase shifts with two distinct characters, referred to as Types~A and B. 
Type~A phase shifts are attributed to coupling of the external magnetic field with an internal Rashba spin-orbit field, resulting in a $\varphi_{0}$-junction. Highly transmissive bound states were shown to make a significant contribution to the phase shift, which was much larger than expected for a single ballistic channel. The discrepancy might be due to the presence of many transverse modes, which future studies could investigate by varying the width and length of the Josephson junction. Type~B shifts were consistent with a $0-\pi$ transition, where orbital effects in the superconducting leads played a critical role. This suggests that the geometry of the superconducting leads, and their impact on orbital effects, is a key ingredient for realizing $\pi$-junctions for superconducting electronics~\cite{Terizioglu1998,Ustinov2003} or in interpreting signatures of topological superconductivity~\cite{Pientka2017}. 

\section{Methods}
Devices were fabricated from a hybrid superconducting-semiconducting heterostructure grown by molecular beam epitaxy on a semi-insulating InP (001) substrate. The heterostructure consisted of a step-graded InAlAs buffer, onto which an $\mathrm{In_{0.75}Ga_{0.25}As}$/InAs/$\mathrm{In_{0.75}Ga_{0.25}As}$ quantum well was grown with a termination of two GaAs monolayers. The step-graded metamorphic buffer compensated the lattice mismatch between the InP and InAs, while the GaAs capping layers provided a barrier for In diffusion into the superconducting layer. The $8~\mathrm{nm}$ InAs layer hosted a two-dimensional electron gas (2DEG), buried $13.4~\mathrm{nm}$ below the semiconductor surface, as measured by transmission electron microscopy~\cite{Cheah2023}. A $15~\mathrm{nm}$ layer of Al was deposited onto the semiconductor surface, \textit{in situ} without breaking vacuum in the growth chamber. Measurements of a gated Hall bar in this material showed a peak mobility of $18000~\mathrm{cm^{2}V^{-1}s^{-1}}$ at an electron sheet density of $8\cdot10^{11}~\mathrm{cm^{-2}}$. This gave an electron mean free path of $l_{\mathrm{e}}\gtrsim260~\mathrm{nm}$, implying that all Josephson junctions measured in this work were in the ballistic regime along the length $L$ of the junction.

The first step in patterning superconducting quantum interference devices (SQUIDs) was to isolate each device from its neighbors by etching large mesa structures. This was done by selectively removing the Al layer with Transene type D, followed by a $380~\mathrm{nm}$ chemical etch into the III-V heterostructure using a $220:55:3:3$ solution of $\mathrm{H_{2}O:C_{6}H_{8}O_{7}:H_{3}PO_{4}:H_{2}O_{2}}$. The second step was to pattern the Al device features, by wet etching in Transene type D at $50^{\circ}\mathrm{C}$ for $4~\mathrm{s}$. A dielectric layer of $\mathrm{Al_{2}O_{3}}$ ($3~\mathrm{nm}$) and $\mathrm{HfO_{2}}$ ($15~\mathrm{nm}$) was deposited across the chip by atomic layer deposition, then gate electrodes were defined on top of the dielectric layer by evaporation and lift-off. Fine gate features were defined in a first step consisting of $5~\mathrm{nm}$ Ti and $20~\mathrm{nm}$ Au; a second deposition of Ti ($10~\mathrm{nm}$) and Al ($420~\mathrm{nm}$) connected the gates on top of the mesa structures to bonding pads, which were defined in the same step.

Measurements were performed in a dilution refrigerator with a base temperature at the mixing chamber below $10~\mathrm{mK}$. Magnetic fields were applied using a three-axis vector magnet, nominally oriented perpendicular to the device ($\Bperp$) and in the plane of the device ($\Bpar$, $\Bt$). Magnetic fields applied in the direction parallel to the Rashba spin-orbit field, or equivalently the direction perpendicular to the current flow, are denoted by $\Bpar$. The in-plane field was rotated by 90 degrees to give $\Bt$, perpendicular to the spin-orbit field. 

Measurements of the differential conductance were performed with standard lock-in amplifier techniques. An AC voltage $\Vac=3~\mathrm{\mu V}$ was applied to the contact of the superconducting probe with frequency $311~\mathrm{Hz}$, in addition to a DC source-drain voltage $\Vsd$. The AC current $\Ione$ and DC current $I_{\mathrm{SD}}$ flowing through the probe to ground was measured via a current-to-voltage (I-V) converter. The differential voltage across the tunnel barrier $\Vone$ was measured to give the differential conductance $G\equiv\Ione/\Vone$. The transparency of the tunnel barrier was controlled with the gate voltages $(\VtL,~\VtR)$, which are denoted by $\Vt\equiv\VtL=\VtR$ (symmetric configuration). Measurements were performed in the tunneling regime, where $G\ll G_{0}=2\mathrm{e}^{2}/h$. A constant bias offset of $43~\mathrm{\mu V}$ was subtracted from all datasets, due to a DC offset at the I-V converter. Since the tunnel probe was superconducting, the measured conductance was a convolution of the density of states (DoS) in the probe and the superconductor-normal-superconductor (SNS) junction: $G=G_{\mathrm{Probe}}\ast G_{\mathrm{SNS}}$. This amounted to a shift in $G_{\mathrm{SNS}}$ features by $\pm\mathrm{e}\Delta^{*}$. For elevated in-plane magnetic fields, the superconducting gap in the tunnel probe was softened, leading to a finite DoS at low energy. This enabled measurements of the DoS in the SNS junction using an effectively normal probe, such that the measured conductance was directly proportional to the DoS in the SNS junction~\cite{Suominen2017,Nichele2017}. In addition to conductance peaks at high source-drain bias corresponding to Andreev bound states (ABSs), we can attribute some features in the conductance spectrum to multiple Andreev reflections or to disorder in the tunnel barrier and sub-gap states in the DoS of the tunnel probe~\cite{Su2018}. For tunneling spectroscopy measurements at an in-plane magnetic field, a first calibration measurement was performed at each field-value by sweeping the perpendicular field across a range $>\pm3~\mathrm{mT}$. The position of zero perpendicular field was determined from spectroscopic features, including the size of the superconducting gap, the shape and peak conductance of high-bias features, and the sharpness of spectral lines. Then, each spectroscopic map was taken across $>5$ oscillation periods such that spectral features were consistent over the full range. 

Current-biased measurements were performed on the same device. Both contacts at the superconducting probe were floated, such that no current flowed through the probe. The tunnel barrier gate voltages, which also covered large areas of the superconducting loop, were set to ${\Vt=-1.5~\mathrm{V}}$ to deplete the InAs surrounding the Al features, thereby preventing parallel conduction and forming a well-defined current path. A DC current was applied by symmetrically biasing the SQUID loop, such that the device potential was not raised with respect to the ground. Hence, the nominal voltage applied to gate electrodes was the same as the potential difference between gates and the device. A ramped current signal was applied from a waveform generator at a frequency of $133~\mathrm{Hz}$. The voltage drop $\Vtwo$ across the loop was measured with an oscilloscope. The switching current, the current at which the SQUID transitioned from the superconducting to resistive state, was recorded when $\Vtwo$ exceeded a voltage threshold of less than $15~\%$ of the maximum voltage in the resistive state. This measurement was repeated $32$ times, and the resulting switching current values were averaged to account for stochastic fluctuations in the switching current~\cite{Haxell2023}. Values of switching current reported in this work were averaged between values obtained for positive and negative bias currents $\Idc$.

\section{Associated Content}
Supporting Information is available at [URL].

It includes: details on materials and device fabrication; additional details on Reference Device measurements; extraction of the current phase relation and phase shift from switching current measurements; current phase relation measurements in an in-plane magnetic field transverse to the junction axis, along $\Bt$; discussion of the origin of zero bias peaks in tunneling spectroscopy; additional tunneling spectroscopy measurements as a function of transverse in-plane field $\Bt$, at different top-gate voltages $\Vtg$ and in an additional device with large superconducting lead length $\Lsc$; additional measurements of the Type B phase shift in different devices; and a discussion of the kinetic inductance of the superconducting loop. Supporting Information contains additional references~\cite{Peltonen2011,Chen2011,Suominen2017b,Annunziata2010}.

\section{Acknowledgments}
We are grateful to C.~Bruder, W.~Riess and H.~Riel for helpful discussions. We thank the Cleanroom Operations Team of the Binnig and Rohrer Nanotechnology Center (BRNC) for their help and support. F.~N. acknowledges support from the European Research Council (grant number 804273) and the Swiss National Science Foundation (grant number 200021\_201082).

\section{Data Availability}
The data that support the findings of this study are available upon reasonable request from the corresponding author.

\bibliography{Bibliography}

\newpage
\newcounter{myc} 
\newcounter{myc2} 
\renewcommand{\thefigure}{S.\arabic{myc}}
\renewcommand{\theequation}{S.\arabic{myc2}}

\section{Reference Device}
\setcounter{myc}{1}
\begin{figure}
	\includegraphics[width=\columnwidth]{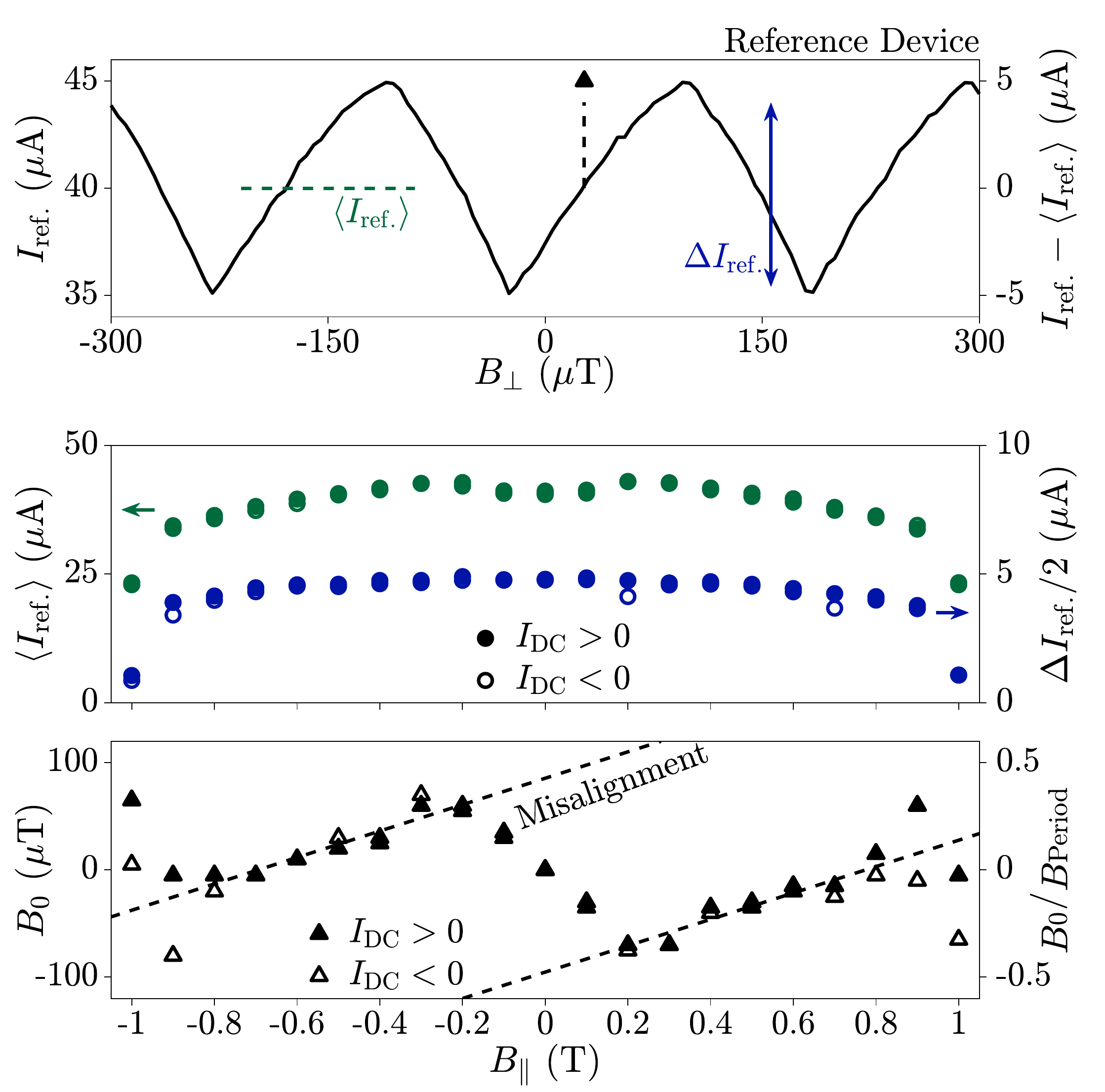}
	\caption{(a) Current-phase relation of Reference Device $\Iref$ before (left) and after (right) subtraction of average $\langle\Iref\rangle$ (indicated by green dashed line). Amplitude of oscillations $\Delta\Iref$ is indicated by the blue arrow. Perpendicular field $\Bperp$ at which $\Iref-\langle\Iref\rangle=0$, $\Bzero$, is indicated by the black triangle. (b) Average switching current of Reference Device $\langle\Iref\rangle$ (green, left axis) and half the oscillation amplitude $\Delta\Iref/2$ (blue, right axis) as a function of in-plane magnetic field $\Bpar$. (c) Perpendicular field offset $\Bzero$ as a function of in-plane magnetic field $\Bpar$ (left axis), and normalized to the oscillation period $\Bperiod$ (right axis). Linear trend in $\Bzero$ for large $|\Bpar|$ (dashed lines) are consistent with residual misalignment of the device chip with respect to the axis of the vector magnet, after appropriate calibration. Values in (b) and (c) are plotted for positive (negative) current bias $\Idc$ as full (empty) markers.}
	\label{Sfig1}
\end{figure}

External magnetic fields were applied using a three-axis vector magnet, nominally aligned in-plane and perpendicular to the surface of the chip. However, small misalignments of the external magnet with respect to the chip mean that large in-plane fields resulted in a perpendicular component, causing a flux through the superconducting loop. To account for this, a Reference Device was fabricated on the same chip, consisting of two Al constrictions in parallel [see Fig.~1(c) of the Main Text]. An example of the switching current of the Reference Device $\Iref$ is shown in Fig.~\ref{Sfig1}(a), as a function of perpendicular magnetic field $\Bperp$. The average switching current $\langle\Iref\rangle=40~\mathrm{\mu A}$ (green dashed line) corresponds to the switching current of the wide Al constriction, $\Wcons=130~\mathrm{nm}$, giving similar values to that of Device 1. The switching current after subtracting the average, $\Iref-\langle\Iref\rangle$ is shown on the right axis. The maximum switching current of the narrow Al constriction, $\Wcons=100~\mathrm{nm}$, is inferred as half the peak-to-peak amplitude of oscillations, $\Delta\Iref/2$. The position where $\Iref-\langle\Iref\rangle=0$ is assumed to be the perpendicular field at which there is no flux threading the loop, $\Bzero$ [marked by the triangle]. 

Figure~\ref{Sfig1}(b) shows the maximum switching current of the wide and narrow constriction as a function of in-plane magnetic field $\Bpar$ (green and blue circles, respectively). Full (empty) markers correspond to the values obtained for positive (negative) applied current $\Idc$. At $\Bpar=0$, the switching current appears to be slightly suppressed relative to that at a small in-plane field. This is attributed to a change in the interplay between quasiparticle populations in the superconductor and the number of quasiparticle relaxation channels in the superconducting leads~\cite{Peltonen2011,Chen2011}. At zero magnetic field, quasiparticles in the Al constriction are confined, with few relaxation channels in the superconducting leads, causing a suppression in the superconducting gap. At small magnetic fields, quasiparticles are generated in the large superconducting leads connected to the constrictions, providing additional relaxation channels for quasiparticles in the constriction region. This partially alleviates the suppression of the superconducting gap relative to the zero-field case, leading to an increase in the switching current. At larger magnetic fields, more quasiparticles are generated in the superconductor resulting in suppression of the switching current. This effect was observed for both in-plane and perpendicular magnetic fields. For $\Bpar>0.9~\mathrm{T}$, a large reduction was observed in the switching current of both constrictions, presumably caused by some portion of the superconducting loop becoming resistive. For this reason, no further studies were performed in this regime. 

The perpendicular magnetic field offset $\Bzero$ of the Reference Device as a function of in-plane magnetic field $\Bpar$ is shown in Fig.~\ref{Sfig1}(c). Misalignment between the vector magnet and the chip is evident at large in-plane magnetic fields, as indicated by the dashed lines. This was considered to be identical for the Reference Device and Device 1 since both are on the same chip. At small $\Bpar$, external fields were distorted, presumably due to flux-focusing effect by the large Al leads~\cite{Suominen2017b}. Flux-focusing effects in the Reference Device for in-plane fields directed along the junction axis, $\Bpar$, were consistent with those measured in all devices.

\section{Extracting the Current-Phase Relation}
\setcounter{myc}{2}
\begin{figure}
	\includegraphics[width=\columnwidth]{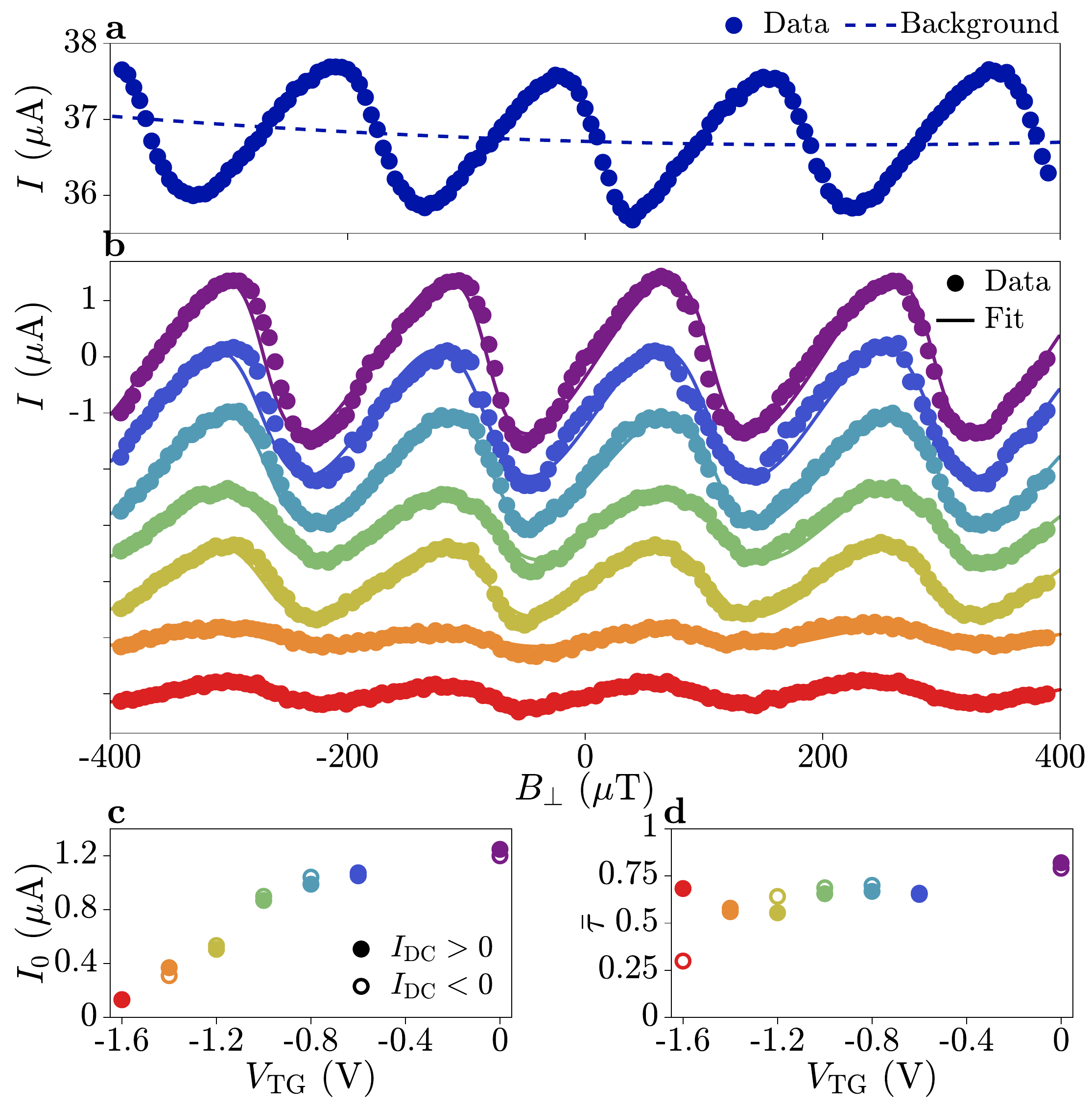}
	\caption{(a) Switching current $I$ of Device 1 as a function of perpendicular magnetic field $\Bperp$. Data (circles) is fitted with a polynomial (dashed line) to extract the background switching current corresponding to the Al constriction. (b) Switching current after background extraction, as a function of perpendicular magnetic field $\Bperp$ for different top-gate voltages $\Vtg$ [colors, defined in (c)]. Data (circles) is fitted with a formula for the current-phase relation of Andreev bound states (line). Each trace is offset by $1~\mathrm{\mu A}$. (c,~d) Results of the fits presented in (b): maximum switching current $I_{\mathrm{C}}$ and transmission $\tau$, for (c) and (d) respectively. Results for positive (negative) applied current $\Idc$ plotted as full (empty) markers.}
	\label{Sfig3}
\end{figure}

An example of the switching current of Device 1 is shown in Fig.~\ref{Sfig3}(a) (circles), as a function of perpendicular magnetic field $\Bperp$. A slowly-varying background is associated with the switching current of the Al constriction, which had a large switching current of $I\approx37~\mathrm{\mu A}$. A weak dependence of the background switching current on $\Bperp$ is consistent with a change in the number and distribution of quasiparticle relaxation channels, as described in the previous section~\cite{Peltonen2011,Chen2011}. To remove this background, the data was fitted with a polynomial function over four complete periods, each defined by $\Bperiod=\mathit{\Phi}_{0}/A=200~\mathrm{\mu T}$ where $\mathit{\Phi}_{0}=h/2\mathrm{e}$ is the superconducting magnetic flux quantum and $A=10.2~(\mathrm{\mu m})^{2}$ is the area enclosed by the superconducting loop. This is shown as the dashed line in Fig.~\ref{Sfig3}(a). Due to the large asymmetry between the critical currents of the SNS junction and the Al constriction, the current-phase relation (CPR) of the SNS junction was taken to be the switching current of the SQUID after subtracting the background. This is plotted as the circles in Fig.~\ref{Sfig3}(b), at $\Bpar=0$ for different top-gate voltages $\Vtg$ [denoted by color, defined in Fig.~\ref{Sfig3}(c)]. The data showed a large forward skewness, consistent with the presence of highly transmissive ABSs in the junction~\cite{Beenakker1991}. 

The CPR of an SNS junction containing $N$ modes is described by
\setcounter{myc2}{1}
\begin{equation}
	I(\varphi) = -\frac{2\mathrm{e}}{\hbar}\sum^{N}_{n=1} \frac{\partial E_{\mathrm{A,}n}(\varphi)}{\partial\varphi},
	\label{eq1}
\end{equation}
where $E_{\mathrm{A,}n}=\mathit{\Delta}\sqrt{1-\tau_{n}\sin^{2}(\varphi/2)}$ is the energy of the $n^{\mathrm{th}}$ ABS with transmission $\tau_{n}$, $\Delta$ is the superconducting gap and $\varphi$ is the phase difference across the SNS junction. The total supercurrent is a sum over the contributions of each ABS in the junction. The junctions studied in this work all had a large width $W=2.5~\mathrm{\mu m}$, and therefore contained many transverse conducting modes. Since detailed knowledge about individual modes is missing, we instead consider an effective transmission $\bar{\tau}$ to describe the properties of the CPR: the transmission which would reproduce the CPR in a junction where all modes have identical transmission. With the application of an in-plane magnetic field, the CPR is expected to obtain a phase shift $\varphi_{0}$~\cite{Yokoyama2014}. Accounting for these considerations, we obtain the equation
\setcounter{myc2}{2}
\begin{equation}
	I(\varphi) = I_{\mathrm{N}}\frac{\bar{\tau}\sin(\varphi-\varphi_{0})}{\Ea(\varphi-\varphi_{0})/\Delta},
	\label{eq2}
\end{equation}
where $I_{\mathrm{N}}=(\mathrm{e}/2\hbar)\bar{N}\Delta$ and $\bar{N}$ is the effective number of modes in the junction. The phase difference across the junction is related to the perpendicular magnetic field by $\varphi=2\pi(\Bperp\cdot A/\mathit{\Phi}_{0})$. The switching current as a function of perpendicular magnetic field is therefore fitted using Eq.~\ref{eq2} obtaining three parameters: $I_{0}$, $\bar{\tau}$ and $\varphi_{0}\equiv 2\pi(\Bzero\cdot A/\mathit{\Phi}_{0})$. The maximum switching current $\Izero$ is not necessarily equal to $I_{\mathrm{N}}$, so it is obtained as the maximum of $I(\varphi)$ from the fit. 

The fits to the data in Fig.~\ref{Sfig3}(b) are shown as the solid lines, with the maximum switching current $\Izero$ and effective transmission $\bar{\tau}$ plotted in Figs.~\ref{Sfig3}(c) and (d), respectively. Note that $\Izero$ is not necessarily equal to the critical current of the SNS junction, since stochastic fluctuations of the phase result in a switching current much lower than the critical current in planar Josephson junctions~\cite{Haxell2023}. The maximum switching current decreased as a function of top-gate voltage $\Vtg$, until no oscillations were visible at $\Vtg<-1.6~\mathrm{V}$. The effective transmission did not change appreciably across this range, indicating the presence of highly transmissive ABSs across the full gate range. Results are plotted for positive ($\Idc>0$) and negative ($\Idc<0$) bias current directions, as the full and empty markers respectively. Changing the current direction resulted in a reversal of the skewness of the CPR, since the external perpendicular field $\Bperp$ had a fixed direction. The sign of the phase $\varphi$ used in Eq.~\ref{eq2} was therefore reversed for negative $\Idc$, as was the associated value of $\Bzero$ coming from the fit. This meant that a larger $\varphi_{0}$ always corresponded to a larger $\Bzero$, independent of the current direction.

\section{Type B Phase Shifts of Current-Phase Relation}
At a given in-plane magnetic field $\Bpar$, the CPR of the SQUID was found by measuring the switching current as a function of perpendicular field $\Bperp$, which was swept multiple times across a small range such that it was stable. The switching current was measured for positive and negative currents, before changing the top-gate voltage $\Vtg$. Once the switching current had been collected for all top-gate voltages, $\Bpar$ was ramped to the next value. The in-plane field was always swept away from $\Bpar=0$, such that sweeps in the positive and negative $\Bpar$ directions began at $\Bpar=0$. As such, all measurements are relative to the values obtained at zero in-plane field in that field sweep. Since fitting with Eq.~\ref{eq2} always returned values for $\varphi_{0}$ in the range $[-\pi,~\pi]$, results at a given in-plane field were shifted by integer multiples of the oscillation period $\Bperiod$ such that $\Bzero$ values followed a monotonic trend. The magnetic field $\Bpar$ was swept multiple times, from $-1~\mathrm{T}$ to $1~\mathrm{T}$, before measurements were taken to minimize hysteresis effects. Nevertheless, some hysteresis was observed at $\Bpar=0$, where flux focusing effects were most prevalent. Hence, results for $\Bpar>0$ and $\Bpar<0$ were combined such that current-averaged $\Bzero$ features were symmetric for $|\Bpar|\geq0.1~\mathrm{T}$. The results of Figs.~1 and 3 of the Main Text were plotted following this procedure. An identical procedure was followed for in-plane magnetic fields applied transverse to the junction axis, $\Bt$.

\setcounter{myc}{3}
\begin{figure}
	\includegraphics[width=\columnwidth]{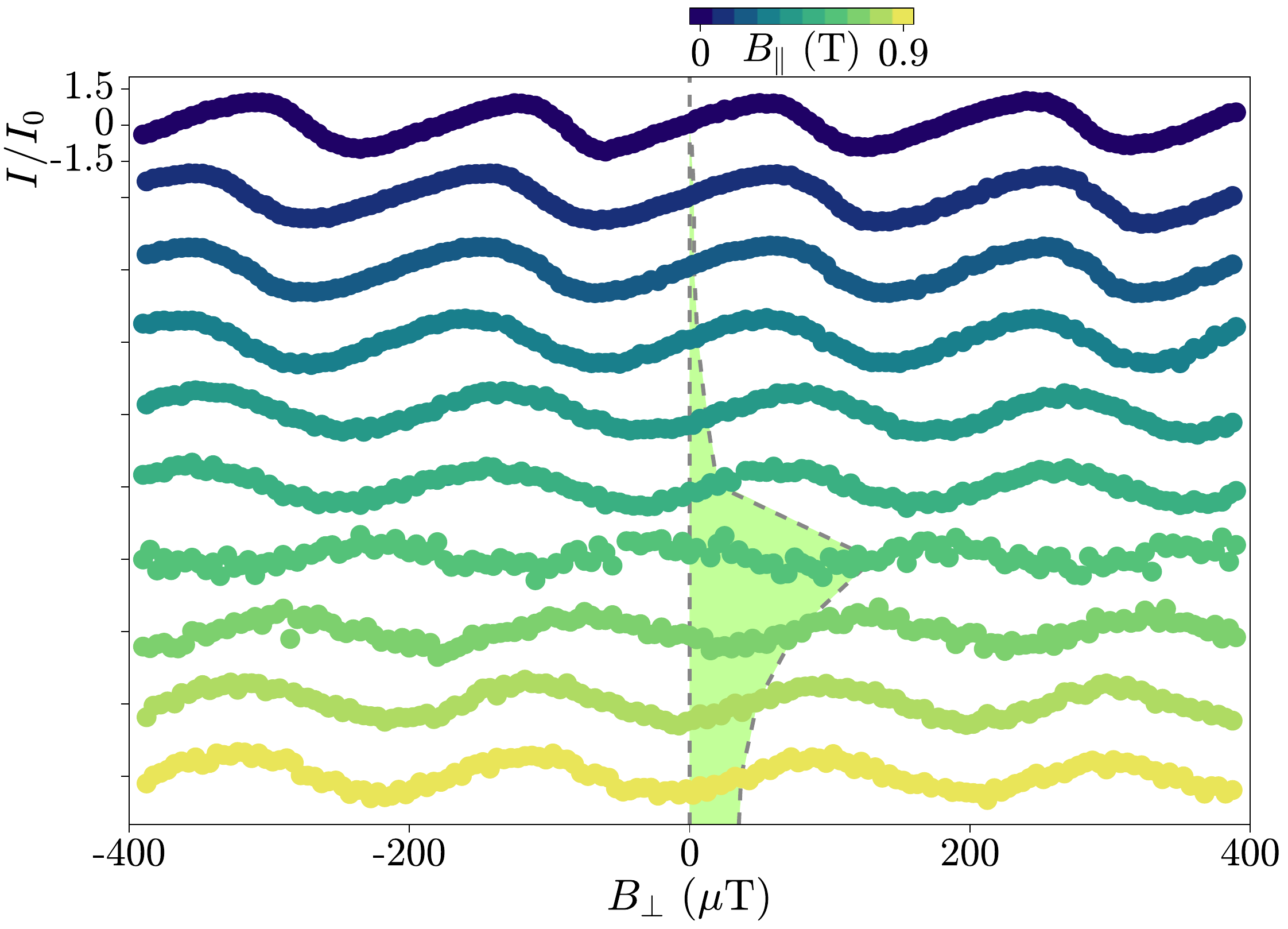}
	\caption{Current-phase relation (CPR) for increasing in-plane magnetic field $\Bpar$. Currents $I$ are normalized to the maximum switching current $\Izero$ at each $\Bpar$. CPR traces are offset by the perpendicular field offset $\Bzero$ of the Reference Device at the corresponding in-plane field $\Bpar$. The shift in $\Bperp$ of the zero-current position is indicated by the green shading, between the two grey dashed lines. Datapoints where the switching current was significantly lower than its neighbors were removed, since they correspond to early switching events in the device by stochastic fluctuations~\cite{Haxell2023}. Each trace is offset by $3~\mathrm{\mu A}$ to improve visibility.}
	\label{Sfig5}
\end{figure}

The CPR as a function of in-plane magnetic field $\Bpar$ is plotted in Fig.~\ref{Sfig5}, where each CPR is normalized to the maximum switching current $\Izero$ at that value of $\Bpar$. The top-gate voltage was $\Vtg=-1~\mathrm{V}$, the same as in the tunneling spectroscopy maps of Fig.~3 in the Main Text. Each CPR trace is plotted with respect to $\Bzero$ of the Reference Device at that in-plane field [see Fig.~\ref{Sfig1}(c)], indicated by the vertical dashed line at $\Bperp=0$. The position of zero current through the SNS junction is marked by the second dashed line, which encloses the shaded green area to $\Bperp=0$. The phase shift shown in Fig.~1(f) of the Main Text is evident, increasing to $\Delta\Bzero/\Bperiod\approx0.5$ at $\Bpar=0.6~\mathrm{T}$, where the switching current is minimal [see Fig.~1(e) of the Main Text]. For larger $\Bpar$, the phase offset moves towards zero, or equivalently towards $\Delta\Bzero/\Bperiod=1$ as shown in Fig.~1(f) of the Main Text. 

This result is consistent with the interpretation of a large phase shift induced by orbital effects in the superconducting leads. As the superconducting gap in the leads is suppressed by orbital effects, ABSs in the junction are pushed closer together, such that some cross zero energy due to Zeeman splitting at the finite in-plane field. When the superconducting gap is sufficiently small, most states have sufficient energy splitting that the ground state is at $\varphi=\pi$ rather than $\varphi=0$~\cite{Yokoyama2014}. This explains the phase shift of $\varphi=2\pi(\Delta\Bzero/\Bperiod)\approx\pi$ at $\Bpar=0.6~\mathrm{T}$, where the orbital effects are strongest. For $\Bpar>0.6~\mathrm{T}$, the superconducting gap in the leads increases as the orbital effects become weaker. This means that fewer ABSs have sufficient energy splitting to shift the phase of the ground state, and $\varphi_{0}$ moves away from $\pi$. The phase shift extends over a range of in-plane fields since the junction contains many ABSs with different transmissions, which will therefore require different Zeeman energies to cross.

\section{Current-Phase Relation Dependence on $\Bt$}
\setcounter{myc}{4}
\begin{figure}
	\includegraphics[width=\columnwidth]{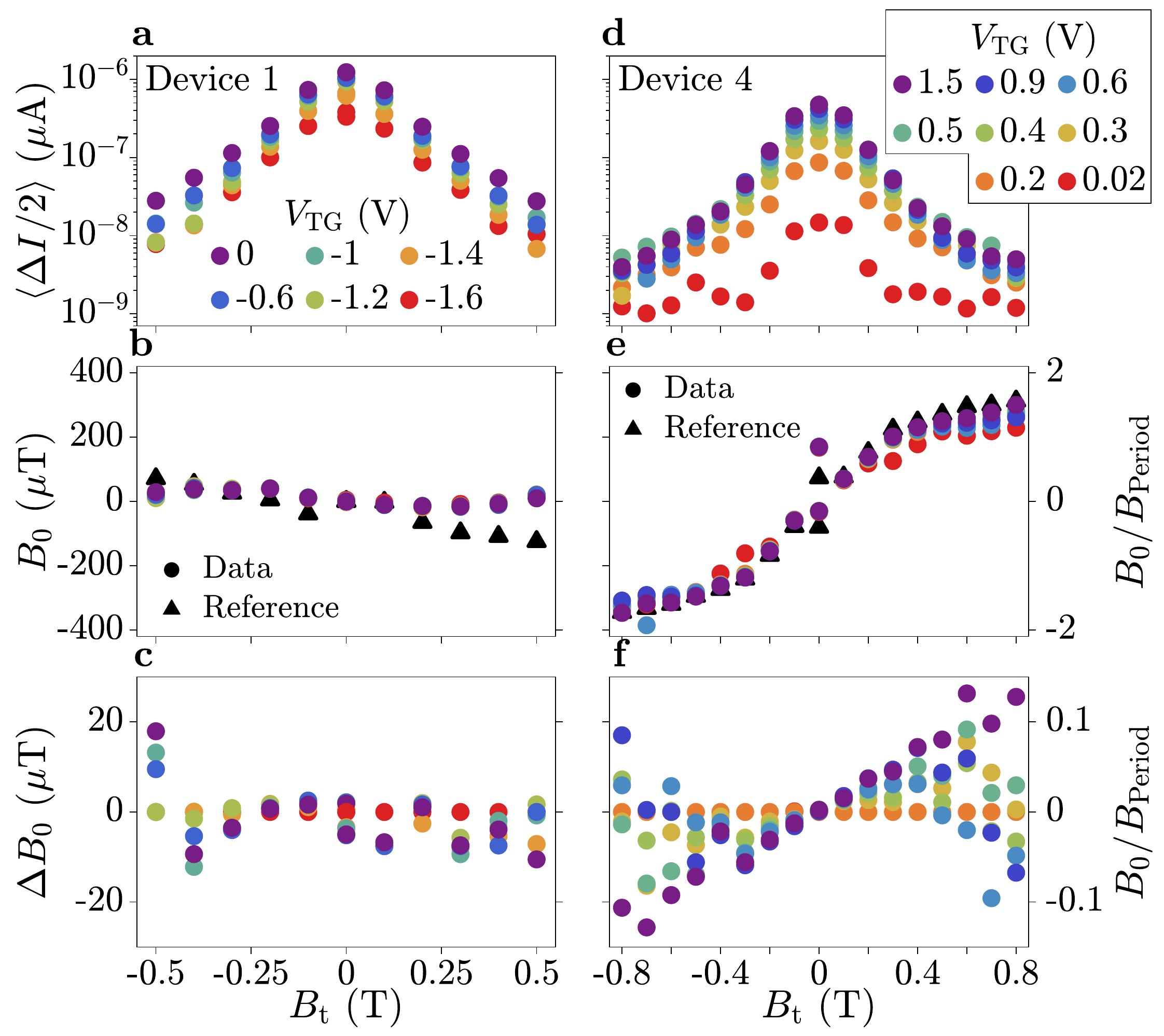}
	\caption{(a) Half-amplitude of switching current oscillations, $\avedI$, for different top-gate voltages $\Vtg$, as a function of in-plane magnetic field $\Bt$. No oscillations in switching current were observed for $|\Bt|>0.5~\mathrm{T}$. (b) Perpendicular field offset $\Bzero$ of switching current oscillations as a function of $\Bt$. The field offset normalized to the oscillation period, $\Bzero/\Bperiod$, is plotted on the right axis. Datapoints for Device 1 (Reference Device) correspond to circles (triangles). The $\Bzero$ for Device 1 and the Reference Device do not align, due to the different level of flux focusing in the two devices. (c) Perpendicular field offset $\Bzero$ plotted with respect to that for the most negative top-gate voltage, $\Vtg=-1.6~\mathrm{V}$. There is no discernable gate-dependent shift across the measured range. (d) Switching current of Device 4 as a function of in-plane magnetic field $\Bt$, for different top-gate voltages $\Vtg$. No $\Bperp$-dependent oscillations in switching current were observed for $|\Bt|>0.8~\mathrm{T}$. (e) Perpendicular magnetic field offset $\Bzero$ in Device 4 (circles) and the associated Reference Device (triangles), as a function of $\Bt$. (f) Field offset $\Bzero$ relative to that at $\Vtg=0.2~\mathrm{V}$, as a function of $\Bt$. Comparison made to $\Vtg=0.2~\mathrm{V}$ rather than the most negative, $\Vtg=0.02~\mathrm{V}$, due to the comparatively large deviation of this datapoint from the Reference Device.}
	\label{Sfig6}
\end{figure}
Phase shifts induced by orbital effects rely on an in-plane mangetic field generating a flux underneath the superconducting leads. This is particular for in-plane fields applied along the junction axis ($\Bpar$), since a field applied in a perpendicular direction ($\Bt$) would not generate desructive interference of ABSs in the superconducting leads~\cite{Pientka2017}. A strong direction dependence is also predicted for spin-orbit related effects, since planar Josephson junctions in InAs are expected to have dominant Rashba spin-orbit coupling directed perpendicular to the junction axis. This has implications for anomalous phase shifts, as well as proposed topological transitions where angular dependence is a crucial ingredient~\cite{Dartiailh2021}. 

Figure~\ref{Sfig6}(a) shows the maximum switching current of SQUID oscillations in Device 1 as a function of in-plane field $\Bt$. The maximum switching current decreased for larger $|\Bt|$, until no oscillations in the switching current were observed for $|\Bt|>0.5~\mathrm{T}$. No minimum and increase in the switching current was observed, nor was there any associated phase jump [Fig.~\ref{Sfig6}(b)], unlike for $\Bpar$ [see Figs.~1(e,~f) of the Main Text]. This is consistent with a lack of orbital effects in the superconducting leads. The small difference between $\Bzero$ for Device 1 and the Reference Device, measured for the same applied $\Bt$, is attributed to different flux focusing effects between the two devices. Figure~\ref{Sfig6}(c) shows the perpendicular field offset relative to the most negative top-gate voltage, $\Vtg=-1.6~\mathrm{V}$. No gate-dependence was present in $\Delta\Bzero$, and there was no linear trend as a function of in-plane field $\Bt$. The absence of gate-dependent phase shifts as a function of $\Bt$ supports the interpretation that Type B phase shifts for $\Bpar$ are enabled by the presence of spin-orbit coupling.

Switching current measurements as a function of $\Bt$ were also performed on Device 4. Figure~\ref{Sfig6}(d) shows the maximum switching current as a function of $\Bt$, for different top-gate voltages $\Vtg$. No minimum and increase in the switching current was observed up to $\Bt=0.8~\mathrm{T}$, beyond which no oscillations in switching current were visible. The corresponding offset in perpendicular field $\Bzero$ [circles, Fig.~\ref{Sfig6}(e)] showed no deviation from that of the Reference Device [triangles, Fig.~\ref{Sfig6}(e)]. This is consistent with Device 1 [Fig.~\ref{Sfig6}], supporting the conclusion that orbital effects do not play a role in measurements in in-plane fields applied perpendicular to the junction axis. Figure~\ref{Sfig6}(f) shows the perpendicular field offset relative to $\Vtg=0.2~\mathrm{V}$. This was chosen to be the reference in this case due to the large deviation of the $\Vtg=0.02~\mathrm{V}$ data from the Reference Device. This was potentially due to the small switching currents at the lowest top-gate voltage, causing an unreliable fit result. Some gate-dependent trend is apparent in Fig.~\ref{Sfig6}(f), although with a smaller gradient than observed for $\Bpar$ [see Fig.~1(f) of the Main Text]. This could be due to stray in-plane fields coupling to the primary spin-orbit direction, or to an additional spin-orbit component in the junction.

\section{Zero-Bias Peak in Tunneling Spectroscopy}
\setcounter{myc}{5}
\begin{figure*}
	\includegraphics[width=\textwidth]{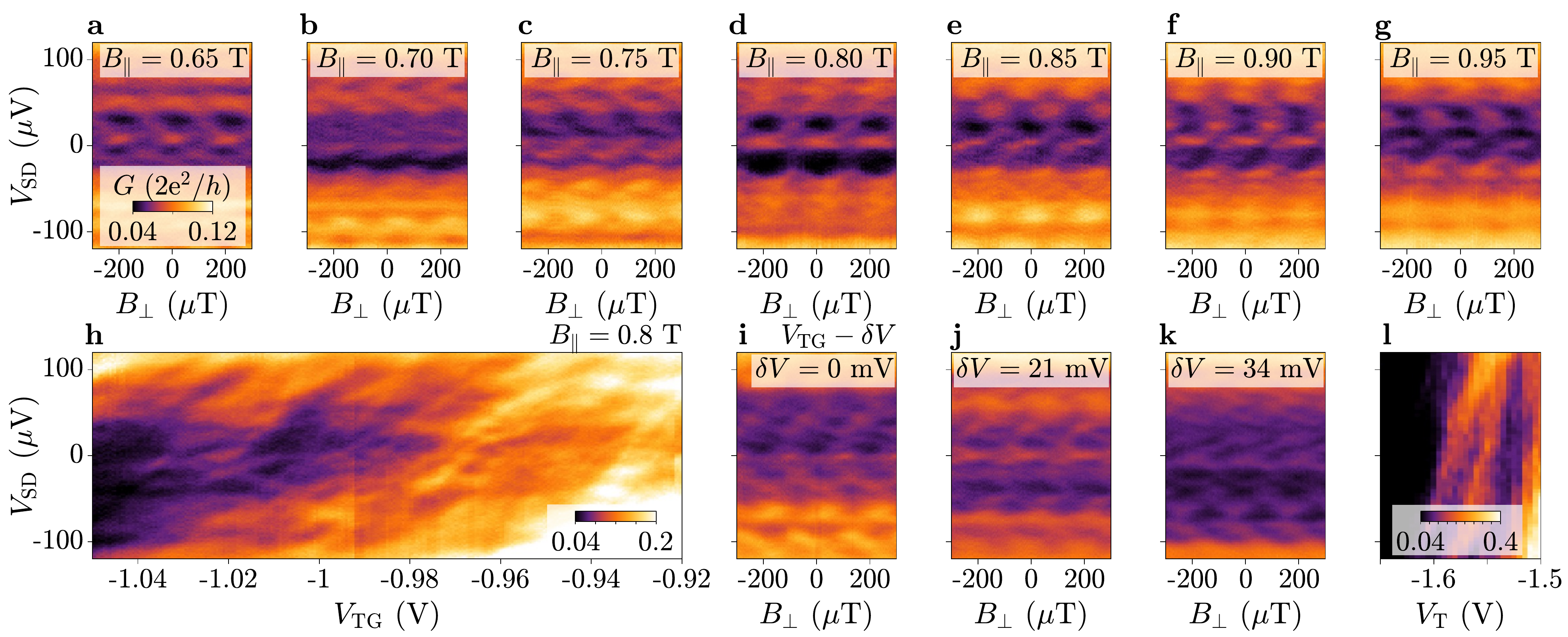}
	\caption{(a-g) Differential conductance $G$ as a function of source-drain bias $\Vsd$ and perpendicular magnetic field $\Bperp$, for different in-plane magnetic fields $\Bpar$. The gate configuration was identical to that of Fig.~3 of the Main Text, with $\Vtg=-1~\mathrm{V}$ [data plotted in (d) is identical to Fig.~3(e) of the Main Text]. (h) Differential conductance as a function of top-gate voltage $\Vtg$, at $\Bpar=0.8~\mathrm{T}$ and $\Bperp=0$. (i,~k) Conductance maps as a function of perpendicular field $\Bperp$, at $\Bpar=0.8~\mathrm{T}$. The top-gate voltage was set to $\Vtg-\delta V$, where $\Vtg=-1~\mathrm{V}$ and $\delta V=0$, $21$ and $34~\mathrm{mV}$ for (i,~k) respectively. (l) Differential conductance as a function of bias $\Vsd$ and tunnel-gate voltage $\Vt$, at $\Bpar=0.8~\mathrm{T}$ and $\Bperp=0$. The top-gate voltage was set to $\Vtg=-1~\mathrm{V}$. High conductance features are tuned by $\Vt$ across the full bias range.}
	\label{Sfig21}
\end{figure*}

Tunneling spectroscopy measurements at large in-plane fields $\Bpar\approx0.8~\mathrm{T}$ show a peak in the differential conductance $G$ close to zero source-drain bias $\Vsd$ [see Fig.~3(e) of the Main Text]. In measurements of similar devices, a zero-bias peak (ZBP) has been associated with the emergence of a topological phase~\cite{Fornieri2019,Ren2019}. Here, we show additional data of the ZBP observed in Fig.~3(e) of the Main Text and comment on its origin. 

Figures~\ref{Sfig21}(a-g) show the conductance $G$ as a function of perpendicular magnetic field $\Bperp$, for in-plane magnetic fields $\Bpar>0.6~\mathrm{T}$ (i.e., after the closure of the superconducting gap at $\Bpar=0.6~\mathrm{T}$). Conductance maps show periodic lobe-like features: each map is plotted such that the center of a lobe is aligned to $\Bperp=0$. The top-gate voltage was set to $\Vtg=-1~\mathrm{V}$, identical to that in Fig.~3 of the Main Text [such that Fig.~\ref{Sfig21}(d) is the same as Fig.~3(e) of the Main Text]. A high-conductance feature is visible close to $\Vsd=0$ in many maps, but does not appear robustly for all in-plane fields and is rarely well separated from conductance features at higher source-drain bias. To test the robustness of this ZBP, the magnetic field was fixed to $\Bpar=0.8~\mathrm{T}$ and $\Bperp=0$, then the top-gate was varied from $\Vtg=-0.92~\mathrm{V}$ to $\Vtg=-1.05~\mathrm{V}$ [Fig.~\ref{Sfig21}(h)]. Conductance features moved close to $\Vsd=0$ as a function of $\Vtg$, but were not stable at $\Vsd=0$ for more than a few millivolts. Figures.~\ref{Sfig21}(i-k) show the differential conductance as a function of perpendicular field $\Bperp$, at top-gate voltages offset from $\Vtg=-1~\mathrm{V}$ by $-\delta V$, where $\delta V = 0$, $21~\mathrm{mV}$ and $34~\mathrm{mV}$ for (i-k) respectively. The conductance spectrum changed appreciably, and a high-conductance feature is evident in Fig.~\ref{Sfig21}(j) but not in the others. Note also that the regime of Fig.~\ref{Sfig21}(d) was not recovered in (i), despite the identical gate and field configuration. Figure~\ref{Sfig21}(l) shows the differential conductance $G$ as a function of tunnel-barrier gate voltage, $\Vt$. High-conductance features were dependent on $\Vt$, and moved across the low-bias region. 

Zero-bias peaks were shown to be sensitive to in-plane mangetic fields $\Bpar$ and top-gate voltage $\Vtg$, and tunnel-barrier-dependent conductance features were shown to move close to $\Vsd=0$. These results suggest that ZBPs were most likely due to ABSs coalescing close to zero energy, rather than being topological in origin. This is despite the gap closure and opening, shown in Fig.~3 of the Main Text and associated with orbital effects in the superconducting leads. This result suggests that additional levels of caution are needed in interpreting ZBPs as indicative of a topological transition, even in the presence of gap closure and reopening. We note that the top-gate voltage $\Vtg=-1~\mathrm{V}$ was chosen to have good visibility of conductance features at low $\Bpar$, to be in a regime of single-subband occupation (based on supercurrent measurements) and to match a value used in supercurrent measurements [see Fig.~1(e-h) in the Main Text]. It was not chosen based on the observation of a ZBP; the emergence of a ZBP after gap closure and reopening was by coincidence rather than by fine-tuning of $\Vtg$.

\section{Tunneling Spectroscopy as Function of $\Bt$}
\setcounter{myc}{6}
\begin{figure}
	\includegraphics[width=\columnwidth]{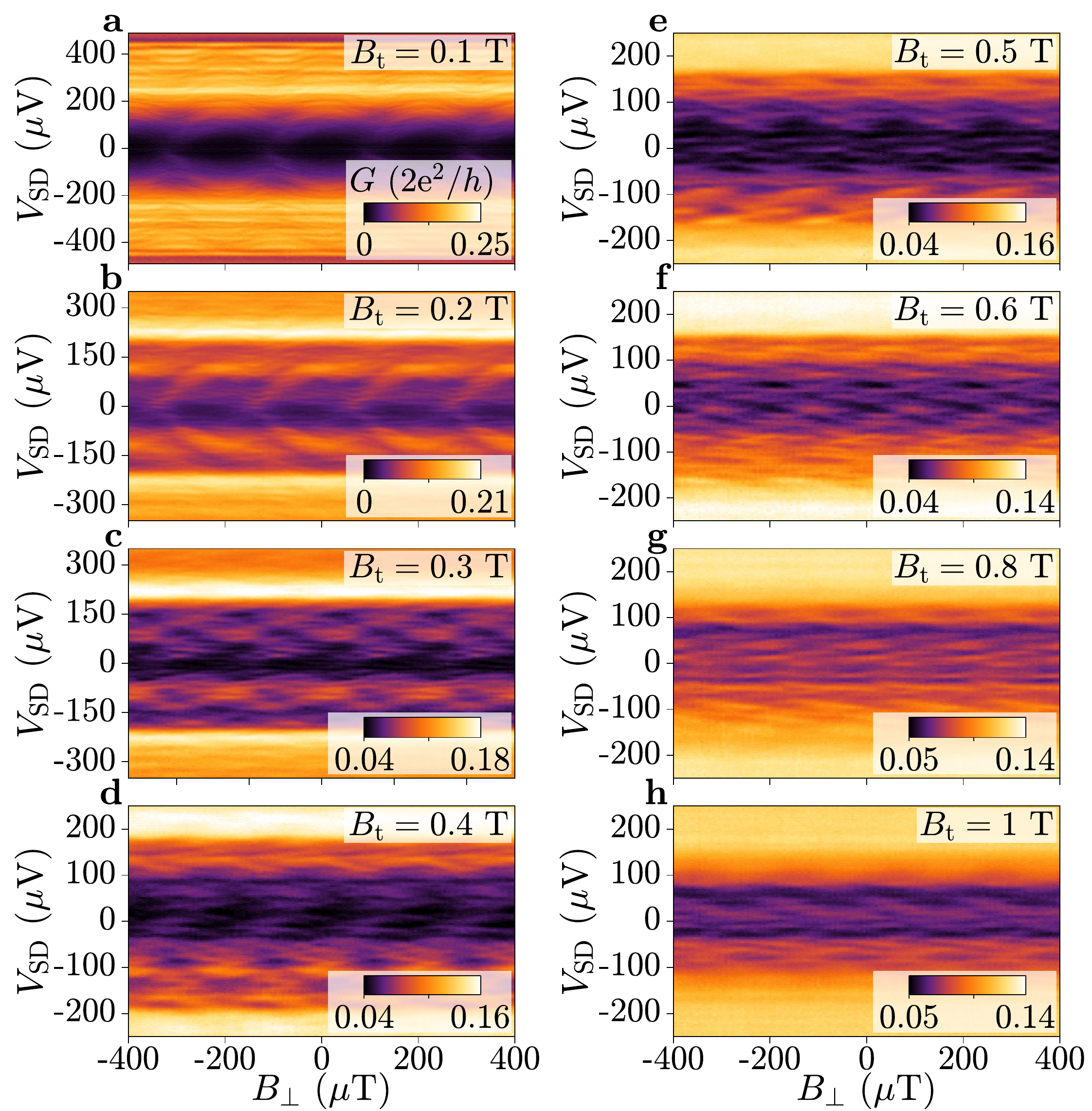}
	\caption{Differential conductance $G$ as a function of source-drain bias voltage $\Vsd$ and perpendicular magnetic field $\Bperp$, for different values of in-plane mangetic field $\Bt$. Measurements were taken at $\Vtg=-1~\mathrm{V}$, in an identical gate configuration as that of Fig.~3 of the Main Text.}
	\label{Sfig7}
\end{figure}

Current-biased measurements for in-plane magnetic fields aligned perpendicular to the junction axis ($\Bt$) are supported by tunneling spectroscopy [see Fig.~\ref{Sfig7}]. Measurements were taken with an identical gate voltage configuration to those in Fig.~3 of the Main Text. For small values of $\Bt$, superconductivity in the tunnel probe was quickly softened such that conductance features occurred at low bias $\Vsd$ [Figs.~\ref{Sfig7}(a,~b)]. Conductance features were periodic with perpendicular magnetic field $\Bperp$, but with a weak dependence consistent with the small switching currents oberved in Fig.~\ref{Sfig6}. Conductance features did not resemble those of ABSs described by $E_{\mathrm{A}}=\mathit{\Delta}\sqrt{1-\tau\sin^{2}(\varphi/2)}$ , instead forming a complex network and crossing $\Vsd=0$ in many places [Figs.~\ref{Sfig7}(c,~d)]. This became more pronounced at larger $\Bt$ [Figs.~\ref{Sfig7}(e,~f)] until the superconducting gap was largely suppressed and conductance features changed very little with $\Bperp$ [Figs.~\ref{Sfig7}(g,~h)]. No reopening of the superconducting gap was observed in these spectroscopic maps, up to large in-plane fields well beyond the value at which no oscillations in the switching current were visible. Conductance features are not well described by a simple model of ballistic ABSs in a short junction, instead showing crossings and interactions at high and low bias. These results indicate the absence of a phase transition, since there was no reopening of the superconducting gap. This is consistent with the lack of orbital effects for in-plane fields applied perpendicular to the junction axis. More sophisticated modeling of ABSs would be required to understand the conductance features in detail, which is beyond the scope of this work.

\section{Tunneling Spectroscopy for different top-gate voltages}
\setcounter{myc}{7}
\begin{figure}
	\includegraphics[width=\columnwidth]{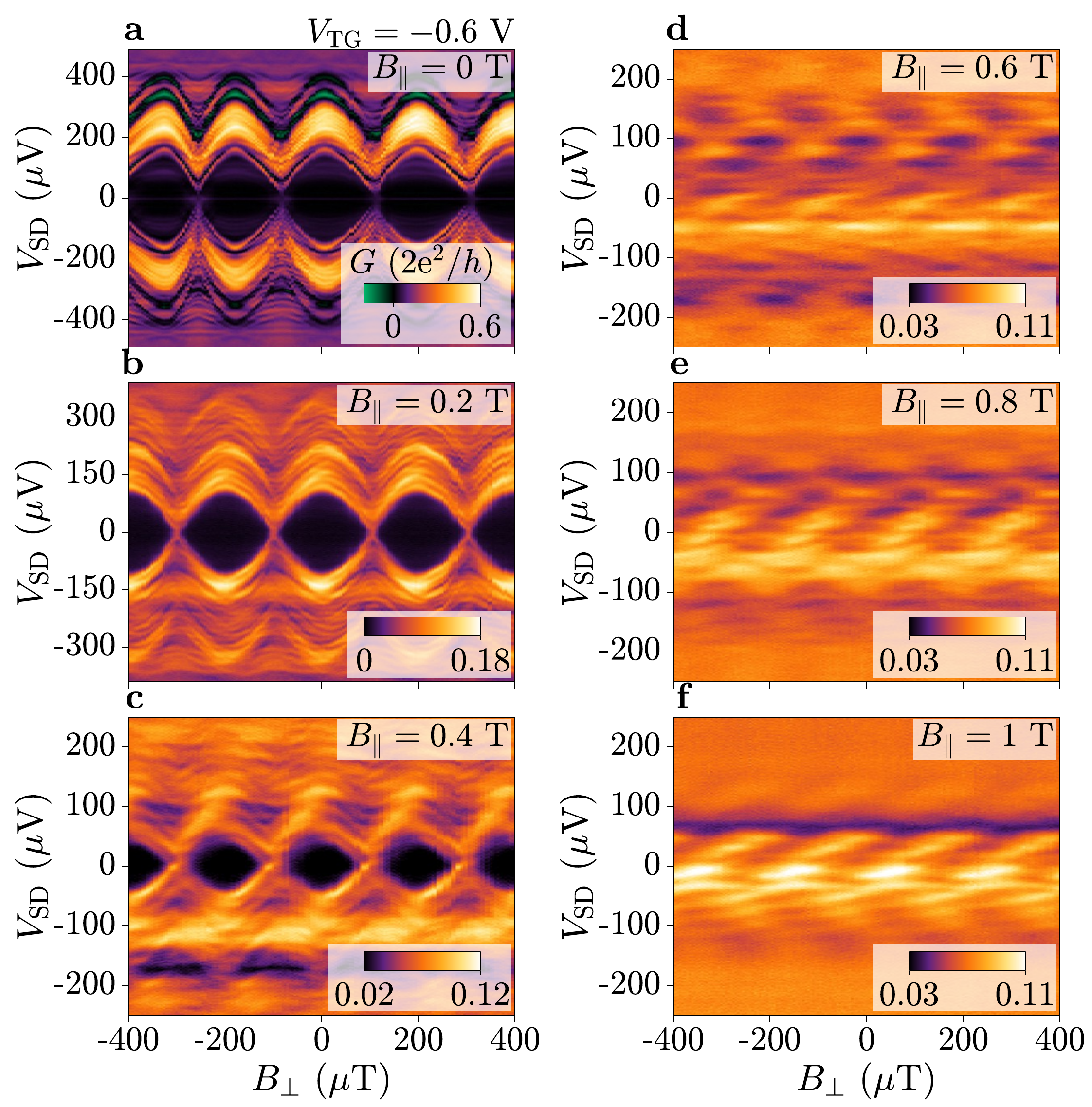}
	\caption{Differential conductance $G$ as a function of source-drain bias $\Vsd$ and perpendicular magnetic field $\Bperp$, for different values of in-plane magnetic field $\Bpar$. Taken at a top-gate voltage of $\Vtg=-0.6~\mathrm{V}$ and tunnel-gate voltage $\Vt=-2.46~\mathrm{V}$.}
	\label{Sfig8}
\end{figure}
\setcounter{myc}{8}
\begin{figure}
	\includegraphics[width=\columnwidth]{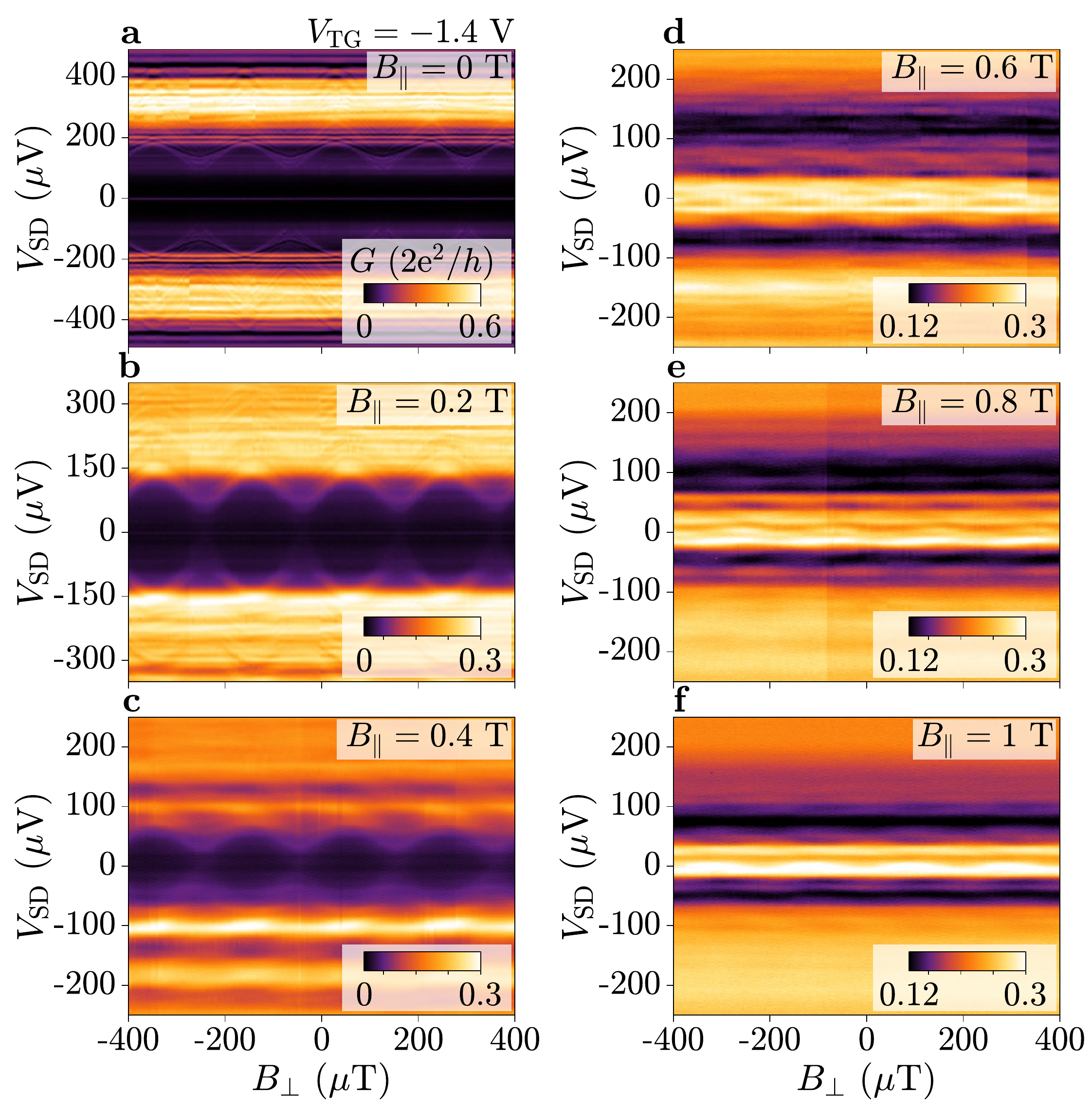}
	\caption{Differential conductance $G$ as a function of source-drain bias $\Vsd$ and perpendicular magnetic field $\Bperp$, for different values of in-plane magnetic field $\Bpar$. Taken at a top-gate voltage of $\Vtg=-1.4~\mathrm{V}$ and tunnel-gate voltages $(\VtL,\VtR)=(-1.835,-1.805)~\mathrm{V}$.}
	\label{Sfig9}
\end{figure}

Figures~\ref{Sfig8} and \ref{Sfig9} show tunneling spectroscopy maps for increasing in-plane magnetic field $\Bpar$, at top-gate voltages of $\Vtg=-0.6~\mathrm{V}$ and $\Vtg=-1.4~\mathrm{V}$ respectively. The tunnel barrier gates were adjusted to be in the tunneling regime, so were set to $\Vt=-2.46~\mathrm{V}$ and $(\VtL,\VtR)=(-1.835,-1.805)~\mathrm{V}$ for Figs.~\ref{Sfig8} and \ref{Sfig9} respectively. At $\Vtg=-0.6~\mathrm{V}$, many more conductance features were present relative to $\Vtg=-1~\mathrm{V}$ [Fig.~\ref{Sfig8}(a) compared with Fig.~3(a) of the Main Text], consistent with more modes present in the junction. In contrast, only few modes were visible at ${\Vtg=-1.4~\mathrm{V}}$ [Fig.~\ref{Sfig9}(a)]. No $\Bperp$-dependent conductance features were observed for top-gate voltages $\Vtg<-1.4~\mathrm{V}$. For increasing in-plane magnetic field $\Bpar$, superconductivity in the tunnel probe was suppressed [Figs.~\ref{Sfig8}(b) and \ref{Sfig9}(b)] and $\Bperp$-dependent conductance features moved closer to $\Vsd=0$ [Figs.~\ref{Sfig8}(c) and \ref{Sfig9}(c)]. At $\Bpar=0.6~\mathrm{T}$, the superconducting gap was suppressed at both top-gate voltages and conductance features had very weak $\Bperp$-dependence close to $\Vsd=0$ [Figs.~\ref{Sfig8}(d) and \ref{Sfig9}(d)]. For larger in-plane fields, some phase-dependence appeared to recover although this was difficult to distinguish due to the poor visibility of conductance features corresponding to individual ABSs [Figs.~\ref{Sfig8}(e,~f) and \ref{Sfig9}(e,~f)]. 

The superconducting gap was suppressed at $\Bpar=0.6~\mathrm{T}$ at all measured top-gate voltages. This is consistent with current-biased measurements [see Fig.~1(e) of the Main Text], where the minimum in the switching current occurred at $\Bpar=0.6~\mathrm{T}$ independent of top-gate voltage $\Vtg$. These results suggest that the cause of gap closure is independent of the properties of the normal region of the junction. Since orbital effects depend only on the properties of the superconducting leads, these findings are consistent with gap closure induced by orbital effects. 

\section{Tunneling Spectroscopy in Device 5}
\setcounter{myc}{9}
\begin{figure*}
	\includegraphics[width=\textwidth]{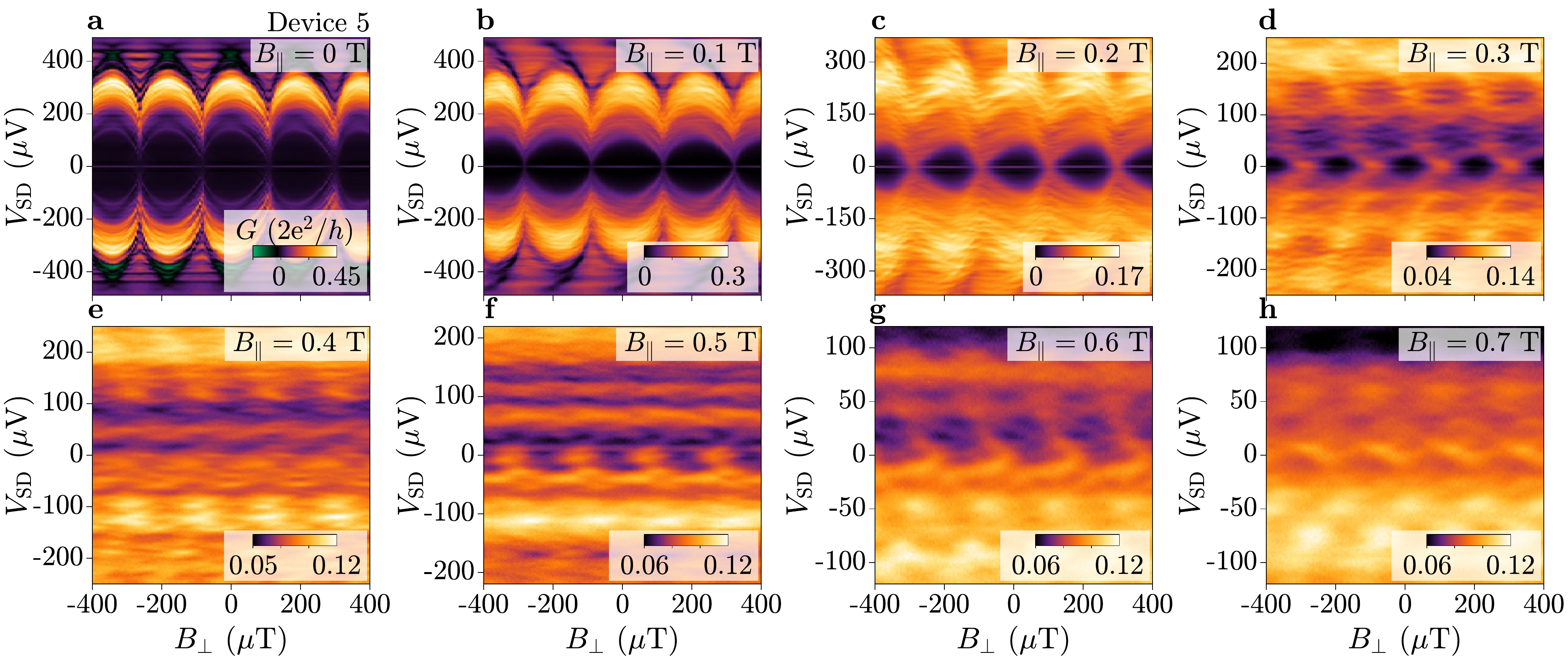}
	\caption{Differential conducance $G$ of Device 5, which was identical to Device 1 other than the superconducting lead length, which was $L_{\mathrm{SC}}=400~\mathrm{nm}$. Conductance maps for different in-plane magnetic fields $\Bpar$, taken at a top-gate voltage $\Vtg=0.8~\mathrm{V}$.}
	\label{Sfig10b}
\end{figure*}
\setcounter{myc}{10}
\begin{figure*}
	\includegraphics[width=\textwidth]{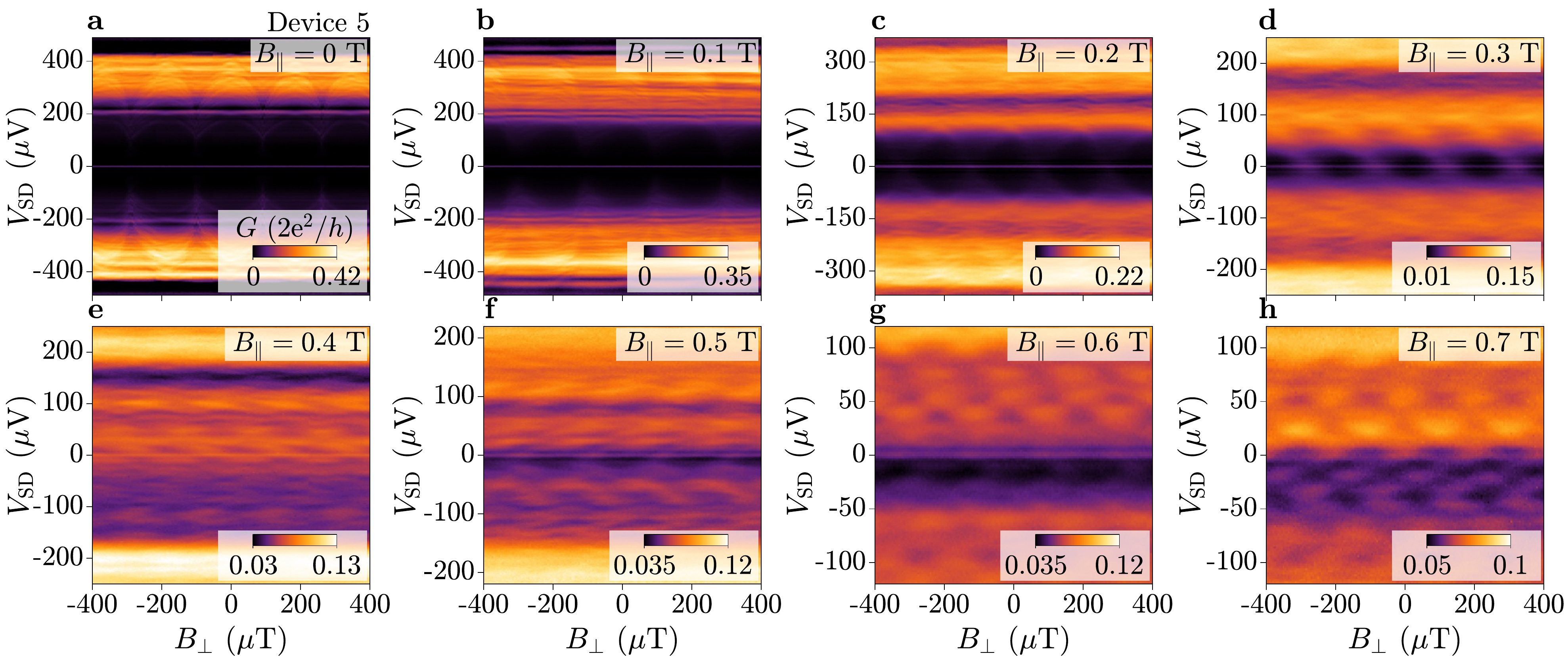}
	\caption{Differential conducance $G$ of Device 5 at different in-plane magnetic fields $\Bpar$, for $\Vtg=0.2~\mathrm{V}$.}
	\label{Sfig10a}
\end{figure*}

Tunneling spectroscopy was performed in an additional device to those shown in the Main Text, which was identical to Device 1 in all aspects other than the length of the superconducting leads $\Lsc=400~\mathrm{nm}$. The superconducting loop in this device, Device 5, was identical to that of Device 2 [Figs.~3(a,~b) of the Main Text], where the switching current was measured. Conductance maps for different values of in-plane magnetic field $\Bpar$ are shown in Figs.~\ref{Sfig10b} and \ref{Sfig10a}, for $\Vtg=0.8~\mathrm{V}$ and $\Vtg=0.2~\mathrm{V}$ respectively. These each correspond to the situation of a large [Fig.~\ref{Sfig10b}(a)] or small [Fig.~\ref{Sfig10a}(a)] number of modes, similar to Figs.~\ref{Sfig8} and \ref{Sfig9} for Device 1. On increasing $\Bpar$, the superconducting gap in the tunnel probe was softened [Figs.~\ref{Sfig10b}(b) and \ref{Sfig10a}(b)] and conductance features moved closer to $\Vsd=0$ [Figs.~\ref{Sfig10b}(c,~d) and \ref{Sfig10a}(c,~d)] until the gap between conductance features was closed at $\Bpar=0.4~\mathrm{T}$ [Figs.~\ref{Sfig10b}(e) and \ref{Sfig10a}(e)]. For larger $\Bpar$, the gap between conductance features reopened and there was a stronger $\Bperp$-dependence [Figs.~\ref{Sfig10b}(f,~g) and \ref{Sfig10a}(f,~g)]. At $\Bpar=0.7~\mathrm{T}$, the gap closed again and superconducting features were suppressed [Figs.~\ref{Sfig10b}(h) and \ref{Sfig10a}(h)]. 

Closure of the superconducting gap was shown to occur at $\Bpar=0.4~\mathrm{T}$ in Device 5, for two top-gate voltages. This is consistent with the minimum in the switching current of Device 2, which had an identical SQUID loop, Al constriction and SNS junction. Tunneling spectroscopy showed a reopening of the gap between conductance features at larger in-plane fields, where a reentrant supercurrent was measured in current-biased experiments. The closure of the superconducting gap and minimum in the switching current both occurred at $\Bpar\approx0.4~\mathrm{T}$, the expected in-plane field at which one flux quantum threads the area underneath the superconducting leads. This supports the conclusion that gap closure in these devices is induced by orbital effects in the superconducting leads.

\section{Devices with Varying Superconducting Lead Length}
\setcounter{myc}{11}
\begin{figure}
	\includegraphics[width=\columnwidth]{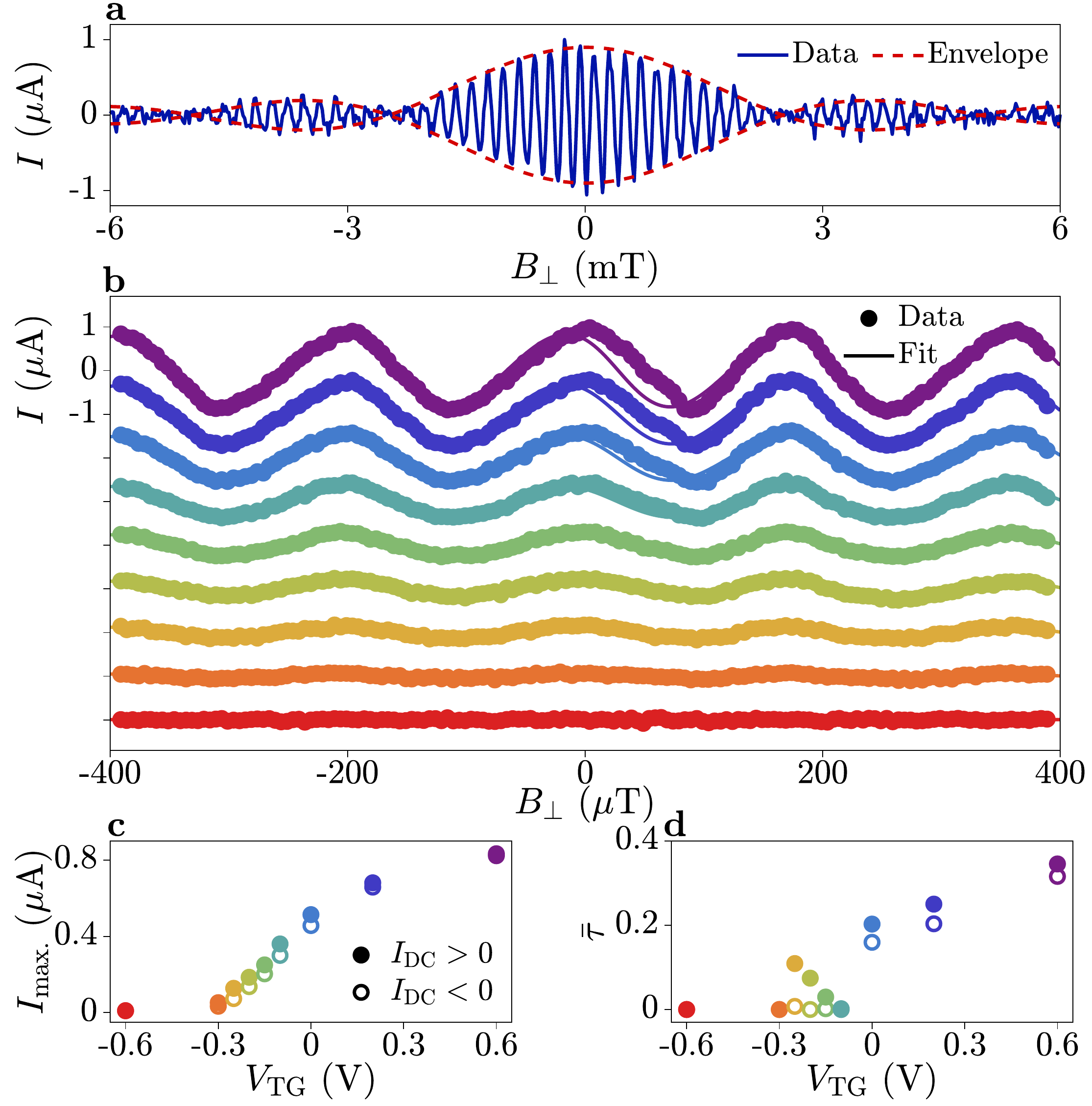}
	\caption{(a) Switching current $I$ of Device 2 as a function of perpendicular magnetic field $\Bperp$, across a wide range of $\pm6~\mathrm{mT}$. Data (blue solid line) is fitted with an envelope function (red dashed line) of a Fraunhofer interference pattern. (b) Switching current as a function of perpendicular magnetic field $\Bperp$, after subtracting the background corresponding to the Al constriction. The background is determined from $\Vtg=-0.6~\mathrm{V}$ (red circles), where the planar junction is considered to be completely closed since no oscillations in switching current were observed. Data at different top-gate voltages (circles) are fitted with a formula for the current-phase relation of Andreev bound states (line), at each top-gate voltage $\Vtg$ denoted by the color [defined in (c)]. The fit incorporates the results obtained for the envelope in (a). Each trace is offset by $1~\mathrm{\mu A}$. (c,~d) Results of the fit presented in (b): maximum switching current $I_{\mathrm{C}}$ and transmission $\bar{\tau}$, for (c) and (d) respectively. Results for positive (negative) applied current $\Idc$ plotted as full (empty) markers.}
	\label{Sfig19}
\end{figure}

Measurements were performed on devices with varying superconducting lead length $\Lsc$ [see Fig.~2 of the Main Text]. Devices consisted of a superconducting loop identical to that of Device 1, other than the length of the superconducting lead which had values $\Lsc=400~\mathrm{nm}$, $350~\mathrm{nm}$ and $180~\mathrm{nm}$ for Devices 2-4 respectively. These devices did not have a tunnel probe proximal to the SNS junction, so only current-biased measurements were possible. Each device had two gates: a top-gate $\Vtg$ identical to that of Device 1 to tune the charge density in the SNS junction; and a global gate covering the exposed InAs regions around the junction and superconducting loop. The global gate was set to $\Vglob<-1.5~\mathrm{V}$ throughout the experiment, such that the exposed InAs was depleted everywhere other than in the junction region.

Switching current measurements were performed for increasing in-plane magnetic field $\Bpar$. At each value of $\Bpar$, the switching current was first measured across a wide range of $\Bperp$ at the most positive top-gate voltage. After subtracting a slowly varying background corresponding to the Al constriction, a recognisable Fraunhofer interference pattern was observed [Fig.~\ref{Sfig19}(a), blue line]. In Devices 2 and 3, where the superconducting leads were large, flux focusing effects were strong. This caused a minimum in the Fraunhofer interference pattern at relatively small perpendicular fields $\Bperp$. It was therefore important to consider the envelope of switching current oscillations due to Fraunhofer interference. This was extracted from the data by filtering out the high frequency oscillatory component, and fitting the result with the following equation

\setcounter{myc2}{3}
\begin{equation}
	I(\Bperp) = I_{0}\left|\mathrm{sinc}\left(\frac{\Bperp-B^{\mathrm{(env)}}_{0}}{B_{\mathrm{min.}}}\right)\right|
	\label{eq3}
\end{equation}
There were three free parameters: the maximum current $I_{0}$, the perpendicular field at which the current was maximum $B^{\mathrm{(env)}}_{0}$ and the perpendicular field at which the first minimum occurred $B_{\mathrm{min.}}$. The result of this fit for the data in Fig.~\ref{Sfig19}(a) is shown as the dashed red line. The in-plane field was aligned such that the maximum of the Fraunhofer pattern was close to $\Bperp=0$ for each value of in-plane field. This was different in each device, due to flux focusing effects, so a different alignment was needed for each device. As such, the Reference Device was measured with each field alignment, to make a direct comparison. 

At a given in-plane magnetic field, the switching current was measured as a function of perpendicular magnetic field $\Bperp$ for different top-gate voltages $\Vtg$. The most negative top-gate voltage was chosen such that no oscillations were visible, where the SNS junction is assumed to be completely closed. The bias current therefore only flowed through the Al constriction, giving a direct evaluation of the switching current of the constriction as a function of $\Bperp$. This background switching current was subtracted from the data at other $\Vtg$, to obtain the current-phase relation at each top-gate voltage [see Fig.~\ref{Sfig19}(b)]. The data (circles) for each $\Vtg$ [colors, defined in (c)] was fitted with Eq.~\ref{eq2}, adjusted to account for the envelope given by Eq.~\ref{eq3}:

\setcounter{myc2}{4}
\begin{equation}
	I(\Bperp) = I_{0}\left|\mathrm{sinc}\left(\frac{\Bperp-B^{\mathrm{(env)}}_{0}}{B_{\mathrm{min.}}}\right)\right|\cdot
	\frac{\bar{\tau}\sin\left[2\pi\frac{(\Bperp-\Bzero)A}{\mathit{\Phi}_{0}}\right]}{\Ea\left[2\pi\frac{(\Bperp-\Bzero)A}{\mathit{\Phi}_{0}}\right]/\Delta}
	\label{eq4}
\end{equation}
Equation~\ref{eq4} takes the fixed parameters $B^{\mathrm{(env)}}_{0}$ and $B_{\mathrm{min.}}$ obtained from the fit to Eq.~\ref{eq3}. There are therefore only three free parameters, as in Eq.~\ref{eq2}: $I_{0}$, $\bar{\tau}$ and $\Bzero$. As for Device 1, $\Izero$ is calculated as the maximum $I(\Bperp)$. The fit for the data in Fig.~\ref{Sfig19}(b) is shown as the colored lines, with the results for $\Izero$ and $\bar{\tau}$ in (c) and (d) respectively [positive (negative) bias currents are indicated by the full (empty) markers]. This procedure is applied to every switching current measurement for Devices 2-4, to obtain the values shown in Fig.~2 of the Main Text. Measurements for positive and negative $\Bpar$ are combined using the same method as for Device 1, as described above. 

\section{Type A Phase Shifts in the Current Phase Relation}
\setcounter{myc}{12}
\begin{figure*}
	\includegraphics[width=\textwidth]{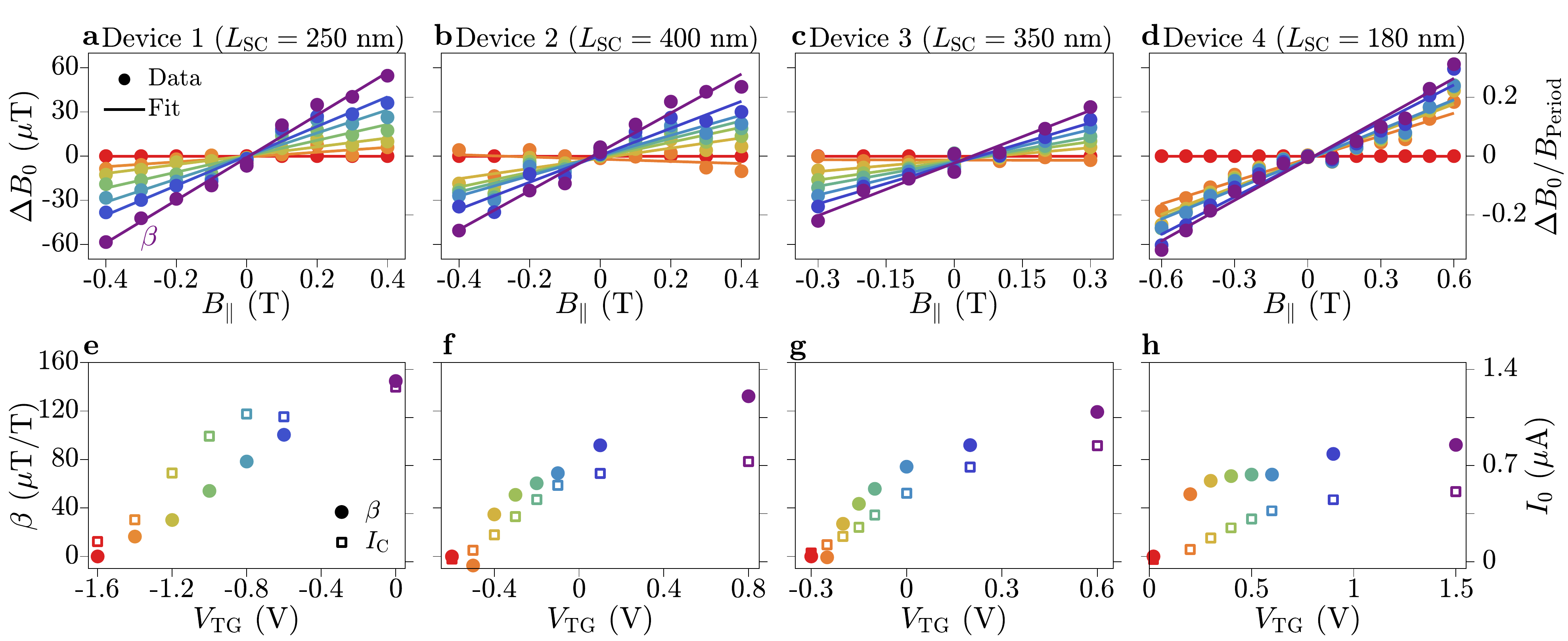}
	\caption{(a-d) Perpendicular field offset $\Delta\Bzero$ relative to the most negative top-gate voltage, as a function of in-plane magnetic field $\Bpar$ for different top-gate voltages $\Vtg$ [indicated by the color, defined in (e-h)], for Devices 1-4 respectively. Data (circles) is fitted with a linear curve at each $\Vtg$ (lines), giving the gradient $\beta$. (e-h) Gradient $\beta$ extracted from (a-d) plotted as a function of top-gate voltage (filled circles, left axis). The maximum switching current as a function of top-gate voltage is also plotted for each Device (empty squares, right axis).}
	\label{Sfig13}
\end{figure*}

Gate-dependent Type~A phase shifts were observed in all devices, for in-plane fields $|\Bpar|\lesssim|\Bphi|$, where $\Bphi$ is the field at which the superconducting gap is suppressed by orbital effects. The results for Devices 1-4 are summarized in Fig.~\ref{Sfig13}. The perpendicular field offset relative to the most negative gate voltage, $\Delta\Bzero$, was linear with in-plane field $\Bpar$ with steeper gradient $\beta$ for more positive top-gate voltage $\Vtg$ [Figs.~\ref{Sfig13}(a-d), colors defined in (e-h)]. The data (circles) are fitted with a linear curve (lines) to extract the gradient $\beta$, which is plotted in Figs.~\ref{Sfig13}(e-h) (filled circles) for Devices 1-4 respectively. The maximum switching current $\Izero$ at $\Bpar=0$ is also plotted as a function of top-gate voltage $\Vtg$ (empty squares). The trend of $\beta$ with $\Vtg$ is similar to that of the maximum switching current $\Izero$.

At the maximum $\Vtg$, where $\Izero$ was large, $\beta\gtrsim100~\mathrm{\mu T/T}$ for all devices independent of the superconducting lead length $\Lsc$. The size of the shift $\Delta\Bzero$ did not depend strongly on the switching current at that in-plane field, rather on the switching current at $\Bpar=0$. This is because the switching current at an in-plane field is significantly influenced by orbital effects, independent of the carrier density at that top-gate voltage. The maximum switching current is linked to the carrier density in the junction, since at lower densities there are fewer transverse modes to carry the supercurrent~\cite{Kjaergaard2017}. The switching current is therefore indicative of the carrier density in the InAs, despite that the gate voltages might differ between devices due to local disorder, inhomogeneous material properties and fabrication imperfections. For decreasing $\Vtg$, the carrier density decreases causing both $\Izero$ and $\beta$ to decrease [Figs.~\ref{Sfig13}(e-h)]. This follows a trend consistent with that of Ref.~\cite{Wickramasinghe2018}, which directly measured the spin-orbit coupling strength as a function of carrier density, in similar InAs quantum wells. 

However, the size of these Type~A phase shifts is much larger than would be expected for a single ballistic channel~\cite{Buzdin2008,Yokoyama2014}, using the spin-orbit coupling strength for InAs~\cite{Wickramasinghe2018}. Similar observations were made in Ref.~\cite{Mayer2020}, where anomalous phase shifts were reported for planar Josephson junctions in InAs/Al heterostructures. The anomalous phase shift was shown to be consistent with that of ABSs in tunneling spectroscopy [Fig.~4 of the Main Text], implying that the phase shift was not dominated by low transmission modes but had contributions from all modes in the junction. 

\section{Type A Phase Shifts in Tunneling Spectroscopy}
Figure~4 shows differential conductance maps for different top-gate voltages $\Vtg$. The perpendicular field at which the ABS energy was lowest was taken to be where the partial derivative of the differential conductance with respect to perpendicular field, $\mathrm{\partial}G/\mathrm{\partial}\Bperp$, was zero at a fixed source-drain bias $\Vsd$. The closest conductance feature to $\Vsd=0$ was considered. This procedure was repeated across 5 lobes, for positive and negative bias, and extracted values of $\Bperp$ were shifted by integer multiples of the period $\Bperiod$ to give values within $[-\Bperiod/2,\Bperiod/2]$. A similar procedure was followed by considering the position where the conductance was closest to $\Vsd=0$, which corresponds to $\varphi\approx\pi$. All methods gave a similar trend and similar quantitative values for the phase shift. The data plotted in Fig.~4(i) of the Main Text is the average of all values obtained from these methods, with the error bars giving the standard deviation.

\section{Phase Shifts due to Kinetic Inductance of the Superconducting Loop}
\setcounter{myc}{13}
\begin{figure}
	\includegraphics[width=\columnwidth]{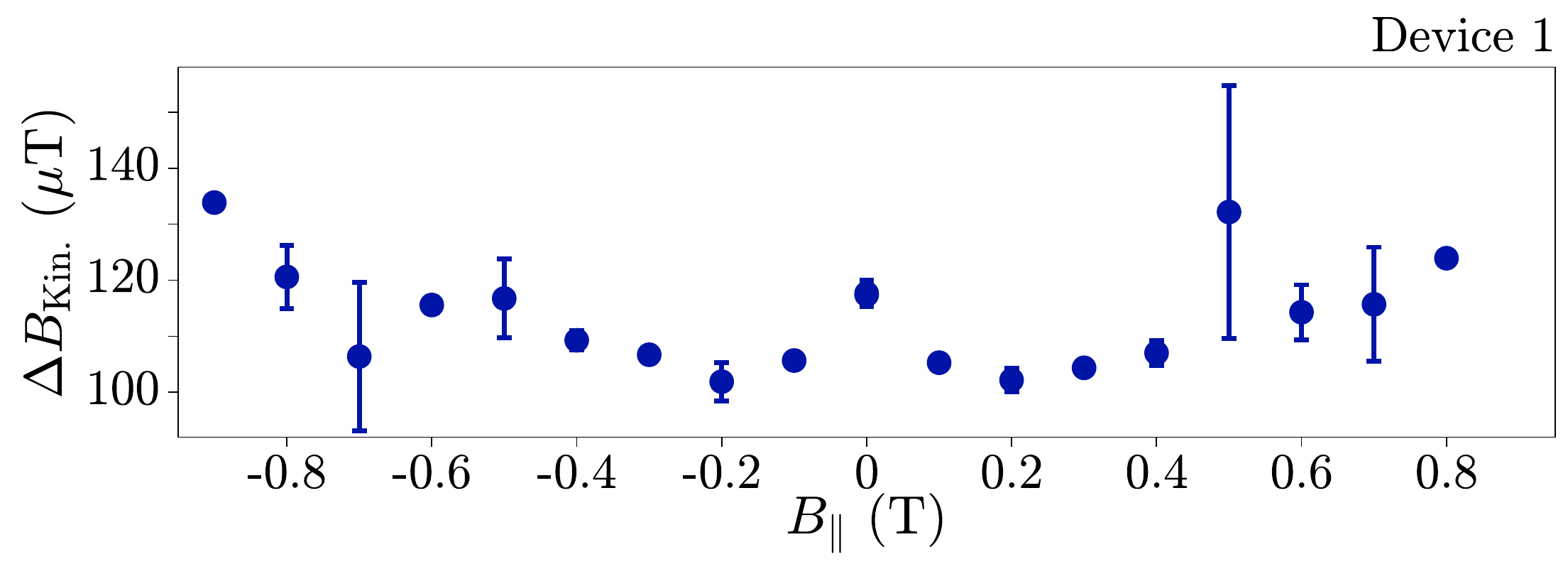}
	\caption{Shift in perpendicular field between current-phase relation traces measured with positive and negative bias currents, $\Delta B_{\mathrm{Kin.}}$. Points are plotted as an average over all top-gate voltages $\Vtg$, with errorbars indicating the standard deviation of all $\Vtg$ values.}
	\label{Sfig22}
\end{figure}
Switching current measurements were performed by applying large bias currents to the SQUID device. Since the epitaxial Al is very thin, it has an appreciable kinetic inductance $L_{\mathrm{K}}$, which generates a flux $\mathit{\Phi}_{\mathrm{K}}=L_{\mathrm{K}}(I_{\mathrm{cons.}}-I_{\mathrm{SNS}})/2$, where $I_{\mathrm{cons.}}$ and $I_{\mathrm{SNS}}$ are the currents flowing in the Al constriction and SNS junction, respectively. The kinetic inductance of the loop is estimated as~\cite{Annunziata2010}
\setcounter{myc2}{5}
\begin{equation}
	L_{\mathrm{K}} = N_{\square}\frac{h}{2\pi^{2}}\frac{R_{\square}}{\Delta}\approx66~\mathrm{pH},
	\label{eq5}
\end{equation}
where $N_{\square}=38$ is the number of squares in the superconducting loop, $R_{\square}\approx1.5~\mathrm{\Omega}$ is the normal-state sheet resistance per unit square measured in a Hall bar geometry on the same material, and $\Delta\approx180~\mathrm{\mu eV}$ is the superconducting gap of Al. This gives a shift of $\Delta B_{\mathrm{Kin.}}\approx110~\mathrm{\mu T}$, for typical currents $(I_{\mathrm{cons.}}-I_{\mathrm{SNS}})$ in the SQUID loop. The shift $\Delta B_{\mathrm{Kin.}}$ between positive and negative currents is shown in Fig.~\ref{Sfig22}. No top-gate dependence was observed, so points were averaged over all top-gate voltages. The field shift $\Delta B_{\mathrm{Kin.}}$ increased for increasing magnitude of in-plane magnetic field, consistent with an increasing kinetic inductance due to quasiparticle generation in the superconducting loop. The values of $\Delta B_{\mathrm{Kin.}}$ in Fig.~\ref{Sfig22} are consistent with the field shift estimated from the kinetic inductance in Eq.~\ref{eq5}.

\end{document}